\documentclass[aps,prd,preprint,groupedaddress]{revtex4-1}
\usepackage{graphicx}
\begin{document}

%mand{\red}{\textcolor{red}}
%\newcommand{\magenta}{\textcolor{magenta}}
\title{\Large Dark Matter Constraints on Low Mass and Weakly Coupled B-L Gauge Boson}

\author{Rabindra N. Mohapatra,$^a$}
\author{Nobuchika Okada$^b$}
\affiliation{$^a$ Maryland Center for Fundamental Physics and Department of Physics, University of Maryland, College Park, Maryland 20742, USA}
\affiliation{$^b$ Department of Physics and Astronomy, University of Alabama, Tuscaloosa, Alabama 35487, USA}

%\date{today}

%\begin{document}

% Use the \preprint command to place your local institutional report
% number in the upper righthand corner of the title page in preprint mode.
% Multiple \preprint commands are allowed.
% Use the 'preprintnumbers' class option to override journal defaults\UTF{0192}%
% to display numbers if necessary
%\preprint{}

%Title of paper

% repeat the \author .. \affiliation  etc. as needed
% \email, \thanks, \homepage, \altaffiliation all apply to the current
% author. Explanatory text should go in the []'s, actual e-mail
% address or url should go in the {}'s for \email and \homepage.
% Please use the appropriate macro foreach each type of information

% \affiliation command applies to all authors since the last
% \affiliation command. The \affiliation command should follow the
% other information
% \affiliation can be followed by \email, \homepage, \thanks as well.
%\author{\bf P. S. B. Dev}
%\affiliation{Dept of Physics,  Wash U.}
%\homepage[]{Your web page}
%\thanks{}
%\affiliation{}

%Collaboration name if desired (requires use of superscriptaddress
%option in \documentclass). \noaffiliation is required (may also be
%used with the \author command).
%\collaboration can be followed by \email, \homepage, \thanks as well.
%\collaboration{}
%\noaffiliation

%\date{\today}

\begin{abstract}
We investigate constraints on the new $B-L$ gauge boson ($Z_{BL}$) mass and  coupling ($g_{BL}$)
  in a $U(1)_{B-L}$ extension of the standard model (SM) with an SM singlet Dirac fermion ($\zeta$) as dark matter (DM).
The DM particle $\zeta$ has an arbitrary $B-L$ charge $Q$ chosen to guarantee its stability. 
We focus on the small $Z_{BL}$ mass and small $g_{BL}$ regions of the model, and find new constraints for the cases 
  where the DM relic abundance arises from thermal freeze-out as well as freeze-in mechanisms. 
In the thermal freeze-out case, the dark matter coupling is given by 
   $g_{\zeta}\equiv g_{BL}Q\simeq 0.016\sqrt{m_\zeta[{\rm GeV}]}$ to reproduce the observed DM relic density
   and $g_{BL}\geq 2.7 \times 10^{-8} \sqrt{m_\zeta[{\rm GeV}]}$ for the DM particle to be in thermal equilibrium prior to freeze-out.  
Combined with the direct dark matter detection constraints and the indirect constraints
   from CMB and AMS-02 measurements, discussed in earlier papers, 
   we find that the allowed mass regions are limited to be $m_\zeta \gtrsim 200$ GeV and $M_{Z_{BL}} \gtrsim 10$ GeV.
We then discuss the lower $g_{BL}$ values where the freeze-in scenario operates and 
   find the following relic density constraints on parameters depending on the $g_{BL}$  range and dark matter mass:
Case (A): for  $g_{BL}\geq 2.7 \times 10^{-8}\sqrt{m_\zeta[{\rm GeV}]}$, 
   one has $g^2_\zeta \, g^2_{BL} + \frac{0.82}{1.2} \, g^4_\zeta \simeq 8.2 \times 10^{-24}$ and
Case (B): for $g_{BL} < 2.7 \times 10^{-8} \sqrt{m_\zeta[{\rm GeV}]}$, there are two separate constraints depending on $m_\zeta$. 
Case (B1): for $m_\zeta \lesssim 2.5$ TeV, we find
$g_\zeta^2 \, g_{BL}^2  \simeq 8.2 \times 10^{-24}  \, \left( \frac{m_\zeta}{2.5 \, {\rm TeV}} \right)$  
   and case (B2): for  $m_\zeta \gtrsim 2.5$ TeV, we have
$g_\zeta^2 \, g_{BL}^2  \simeq 8.2 \times 10^{-24}$. 
For this case, we display the various parameter regions of the model that  can be probed by a variety of ``Lifetime Frontier" experiments
  such as FASER, FASER2, Belle II, SHiP and LDMX. 
%
%Even for the freeze-in case, the direct detection constraints set a lower bound on, for example, $M_{Z_{BL}} \gtrsim 70$ MeV 
%for the DM mass of 30 GeV. 
%
%We also discuss the bounds from SN1987A observations on the $g_{BL}$ analogous to the case of dark photon
%for $M_{Z_{BL}}\leq 100$ MeV.
\end{abstract}
\maketitle

%%%%%%%%%%%%%%
\section{Introduction}  
%%%%%%%%%%%%%%
Extensions of the standard model (SM)  with $U(1)_{B-L}$ as a possible new symmetry of electroweak interactions, is well motivated due to its connections to the neutrino mass ~\cite{marshak1,marshak2} and  has recently attracted a great deal of attention. Theoretical constraints of anomaly cancellation allow two classes of $B-L$ extensions: (i) one motivated by left-right symmetric and SO(10) models, where the $B-L$ generator contributes to the electric charge of particles \cite{marshak1,marshak2,Davidson} 
and (ii) another, where it does not \cite{BL0, khalil1, BL1, BL1a, BL3, BL2, BL4}. 
The second alternative is not embeddable into the left-right or SO(10) models. Both classes of models require the addition of three right handed neutrinos to satisfy the anomaly constraints and lead to the seesaw mechanism for neutrino masses~\cite{seesaw1, seesaw2, seesaw3, seesaw4, seesaw5}. There is however a fundamental difference between the two classes of models as regards the possible magnitudes of their gauge couplings: in the first class of models where the $B-L$ contributes to electric charge~\cite{marshak1,marshak2,Davidson}, there is a relation between the electric charge of the positron and the $B-L$ gauge coupling:
\begin{eqnarray}
\frac{1}{e^2}=\frac{1}{g^2_L}+\frac{1}{g^2_R}+\frac{1}{g^2_{BL}}. 
\end{eqnarray}
As a result, there is a lower bound on the value of $g_{BL}$:  
\begin{eqnarray}
\frac{1}{g^2_{BL}}\leq \frac{\cos^2\theta_W}{e^2} ~~{\rm or} ~~g_{BL} \geq 0.34. 
\end{eqnarray}
This lower bound gets strengthened to $0.416$, when it is assumed that all $U(1)$ couplings in the $SU(2)_L\times U(1)_{I_{3R}}\times U(1)_{B-L}$ model are perturbative till the Grand Unified Theory scale~\cite{garv}.

In the second class of models on the other hand, there is no lower bound on $g_{BL}$ from theoretical considerations, and as a result, it can be arbitrarily small.  In this paper, we focus on this class of models in the small $g_{BL}$ and small $B-L$ gauge boson mass ($M_{Z_{BL}}$) regions to see what kind of phenomenological constraints exist, once we add a Dirac dark matter fermion $\zeta$ to the theory. 
We let the dark matter (DM) field have an arbitrary $B-L$ charge, $Q$. 
Clearly, it is possible to choose a $B-L$ charge $Q$ for $\zeta$ so that it is naturally stable as is required for a dark matter particle. 
For example,  if we choose $Q$ to be a half odd integral value, there are no operators in the theory that will make it decay. 
This class of models are completely realistic as far as the their fermion sector is concerned. 
There are four parameters: $g_{BL}$, $g_{\zeta}\equiv g_{BL}Q$ plus the two mass parameters, $m_{\zeta}$ and $M_{Z_{BL}}$, 
 which enter into our dark matter discussion. 
See Refs.~\cite{FileviezPerez:2019cyn, Gu:2019ohx} for the case where the two mass parameters in the multi-TeV range. 
We keep the masses arbitrary and find constraints on them in our model. 
%%%%%%%%%%%%%%%%%%%%%%%%%%%%
Although our interest is mostly phenomenological in this paper and therefore we do not worry
  about the origin and naturalness of small gauge couplings, 
  we do note that small gauge couplings are motivated by a class of large volume compactification of string theories 
  (see, for example, Ref.~\cite{burgess}).  
%%%%%%%%%%%%%%%%%%%%%%%%%%%%%%%
We also  ignore mixings between the $B-L$ gauge boson and the SM gauge bosons 
   as well as the mixing between $Z_{BL}$ and the photon, for simplicity.   
As a result, there are no mixing effects in the $Z_{BL}$ couplings.  
In any case,  these mixing effects are loop suppressed and therefore smaller than the effects we have considered.
The DM particle, $\zeta$, in our  case is a Dirac fermion, as just mentioned and gauge anomaly cancellation is automatically satisfied. 
To emphasize again, $\zeta$ is stable due to the choice of its $B-L$ charge.

We discuss constraints that $g_{BL}$ and $g_\zeta$ must satisfy from the requirements that the particle $\zeta$ be a viable dark matter 
  i.e.~it satisfies the relic density constraints as well as direct detection constraints and other indirect detection constraints
  such as from cosmic microwave background (CMB) and cosmic ray measurements.  
We consider the following two gauge coupling parameter ranges of the theory: 
(i) one where the DM relic density arises via thermal freeze-out and (ii) the second case where the couplings, 
  $g_{BL}$ and $g_\zeta$, are so small that the DM particle $\zeta$ was never in thermal equilibrium
   in the early universe with SM particles %but $g_{BL}$ was such that it was in thermal equilibrium with the SM fermions 
   and it had a vanishing density at the reheating after inflation.  
The DM relic abundance in the latter case was built up via the freeze-in mechanism~\cite{hall,bernal,hambye,chu}. In the first case, we find that the relic density constraint requires that $g_\zeta \simeq 0.016 \sqrt{m_\zeta[{\rm GeV}]}$ and the condition for thermal equilibrium of $Z_{BL}$ in the early universe requires that $g_{BL}\gtrsim 2.7 \times 10^{-8}\sqrt{m_\zeta[{\rm GeV}]}$. 
%%%%%%%%%%%%%%%%
For the freeze-in case, 
we find that the product $g_{BL} \, g_\zeta \approx 2.9 \times 10^{-12}$  to satisfy the constraint of the DM relic density.
This result is independent of the dark matter mass as long as $m_\zeta \gtrsim 2.5 \, {\rm TeV}\gg M_{Z_{BL}}$. 
When the dark matter mass is less than 2.5 TeV, the so-called sequential freeze-in mechanism dominates
and the condition on couplings becomes $g_{BL} \, g_\zeta \approx 2.9 \times 10^{-12}\sqrt{m_\zeta/2.5 \, {\rm TeV}}$
%for the same DM and $Z_{BL}$ mass values 
%
(the freeze-in mechanism for a Majorana fermion DM and $g_\zeta=g_{BL}$ 
was investigated in Ref.~\cite{Kaneta} and their results are consistent with ours). 
It is interesting that the spin-independent direct detection cross section also depends on the product  
 $(g_{BL} g_\zeta)^2 \mu_{\zeta N}^2/M_{Z_{BL}}^4$ (where $\mu_{\zeta N}$ is the reduced mass of the DM-nucleon system) 
and therefore the $\Omega_{DM}$ constraint also puts lower limits on the $Z_{BL}$ mass. We explain the origin of these constraints and elaborate on the details in the body of the paper.

We next comment on two more cases: Case (iiiA) where the $g_\zeta$ is large enough that both $Z_{BL}$ and $\zeta$ were in equilibrium with each other but not with the SM particles and Case (iiiB) where both $g_{\zeta, BL}$ are so small that all three sectors were thermally sequestered from each other.  
These cases do not fall into either the freeze-in or freeze-out scenarios and are therefore listed separately.
%{\bf Fermi Lat detailed comments here!}

There are also constraints on this model from Fermi-LAT observations that assume 100\% branching ratio to either $b\bar{b}$ or $\tau^+\tau^-$~\cite{fermi} which are compatible with the thermal freeze-out constraints only for $m_\zeta \geq$ few GeV. The assumption of 100\% branching ratio is however not the case for our model and we have more like 20\% for the branching ratio. As a result, our bounds are weaker and we estimate it to be in the 2 GeV range for the freeze-out case using the Fig.~9 of the Fermi-LAT paper~\cite{fermi}.

%%%%%%%%%%%%%%%%%%%%%%%%%%%%%%
We note here that there are  other $B-L$ models with dark matter in the literature \cite{bauer, biswas} as well as $B-L$ models without the dark matter \cite{heeck}. 
There are also models with dark photon \cite{lindner} and dark $U(1)$ models \cite{cirelli} with some similarity 
to $B-L$ models.  %as well as $U(1)$ models with similar properties to our model.  
Our model is however different from all of them. 
For example, Ref.~\cite{heeck} discusses constraints $g_{BL}$ and $M_{Z_{BL}}$ for a pure $B-L$ model with Dirac neutrinos without any dark matter whereas our model not only has a dark matter but also the neutrinos are Majorana particles which obtain their mass from the seesaw mechanism resulting from $B-L$ breaking. 
Furthermore, we consider the case where the $B-L$ gauge boson couples to the dark matter having an arbitrary $B-L$ charge. 
As far as Ref.~\cite{biswas} is concerned, it uses the lightest right handed neutrino as the dark matter and as a result, its $B-L$ charge of DM is fixed by anomaly cancellation. On the other hand, in our model, the dark fermion is separate from the usual SM plus the right handed neutrinos model. As a result, we can choose its $B-L$ charge arbitrary consistent with anomaly cancellation. This allows us to explore a very different range of parameters of the $B-L$ model. Our model is also different from other $U(1)$ based models e.g.~Refs.~\cite{cirelli,lindner}, although they have some similarity to our discussion e.g.~their constraints on dark photon portal models with an MeV dark matter (see Ref.~\cite{lindner}). %This has similarity to some of our considerations. In particular, 
We have used some results from this paper e.g. the CMB bounds on dark matter using Fig.~3 of Ref.~\cite{lindner} 
  which imply the constraint of dark matter mass of $m_\zeta \geq 1$ GeV.  
To be consistent with the bounds, in this paper, we focus on the region of dark matter mass,  $m_\zeta \geq 1$ GeV. 
%%%%%%%%%%%%%%%%%%%%%%%%%%%%%%%%%

%If the $g_{BL}$ coupling is much weaker than $10^{-7}$ given above, the $\zeta$ particle can still be a viable dark matter acquiring its relic density via the so-called freeze-in mechanism~\cite{hall}. 

The paper is organized as follows: in Sec.~\ref{sec:2}, we outline the details of the model.  
In Sec.~\ref{sec:3}, we discuss the case of thermal freeze-out of the dark matter 
  and the constraints on the relevant model parameters from it.  
We then combine it with the already existing indirect detection constraints to find new allowed regions 
  for the DM mass for different $M_{Z_{BL}}$ values. 
In Sec.~\ref{sec:4}, we switch to the parameter range of the model where the relic density arises out
  of the freeze-in mechanism and the constraints implied by it on the model. 
We note how the FASER experiment~\cite{faser1} combined with other planned/proposed experiments
   such as Belle II, SHiP and LDMX can probe parameter range of the model.   
We also comment on constraints from the SN1987A and Big Bang Nucleosynthesis (BBN).  
In Sec.~\ref{sec:5}, we briefly discuss the case where the ``dark sector'' with $\zeta$ and $Z_{BL}$ is decoupled  
   from the SM thermal plasma and are produced from the inflaton decay at the end of inflation. 
We conclude in Sec.~\ref{sec:6} with a discussion of implications of our results and some additional comments.

%%%%%%%%%%%%%%%%%%%%%%%%%%%%%%%
\section{The $B-L$ model with Dirac fermion dark matter} 
\label{sec:2}
%%%%%%%%%%%%%%%%%%%%%%%%%%%%%%%
\subsection{Model details}
Our model is based on the $U(1)_{B-L}$ extension of the SM with gauge quantum numbers under $U(1)_{B-L}$ defined by their baryon or lepton number of particles. The gauge group of the model is $SU(3)_c \times SU(2)_L\times U(1)_Y\times U(1)_{B-L}$, where $Y$ is the SM hypercharge. We need three right handed neutrinos (RHNs) with $B-L=-1$ to cancel the $B-L$ anomaly. 
The RHNs being SM singlets do not contribute to SM anomalies. The electric charge formula in this case is same as in the SM. 
We now add to this model a vector-like SM singlet fermion $\zeta$ with $B-L$ charge equal to $Q$. 
Being vector-like, this fermion does not affect the anomaly cancellation of the model. The $B-L$ group is assumed not to contribute to electric charge formula as stated in the introduction. As a result, its couplings are theoretically not restricted.
We assume that there is a Higgs boson with $B-L=+2$ which gives a Majorana mass to the RHNs thereby helping to implement the seesaw mechanism for neutrino masses since the SM Higgs doublet already provides the Dirac mass to the neutrinos. 
%The dark matter fermion in this case is a Dirac fermion.  
The interaction Lagrangian in our model describing the interaction of the $B-L$ gauge boson (called $Z_{BL}$ here) is:
\begin{eqnarray}
{\cal L}_{Z_{BL}}= \left( Z_{BL} \right)_\mu \left[ g_{BL}\sum_f (B-L)_f \bar{f}\gamma^\mu f
+g_\zeta \bar{\zeta}\gamma^\mu \zeta~\right]. 
%+~h.c.
\end{eqnarray} 
This Lagrangian is enough to derive our conclusions. We start with letting the values of $g_{BL}$, $g_\zeta \equiv Q g_{BL}$, $M_{Z_{BL}}$ and $m_\zeta$ as free parameters and explore the smaller mass range of $M_{Z_{BL}}$ and as a benchmark point, 
  we take $m_\zeta $ in the range of $1$ GeV to few TeV range with $M_{Z_{BL}} < m_{\zeta}$. 
Clearly this covers a wide and interesting range of dark matter masses. 
%The direct detection constraints on $m_\zeta$ vs $g_{BL}$ is shown in ~\cite{mexico}

\subsection{New Higgs bosons and other phenomenology}
The only new Higgs boson in the model beyond the SM Higgs doublet, is an SM singlet field $\Delta$ with $B-L=2$. 
It acquires a non-zero vacuum expectation value $\langle \Delta \rangle= v_{BL}$. The real part of $\Delta$ is a physical Higgs field, 
which we denote by $\sigma$. It couples to the right handed neutrinos which we assume heavy (in the TeV range or higher)
so that $\sigma$ could be a long lived particle. 
Also it has no direct couplings to quarks and leptons and %its only coupling to quarks and leptons 
such couplings arise from its mixing with the SM Higgs boson. 
For a GeV mass $\sigma$, we may expect this mixing to be of order $m_\sigma^2 /m_h^2 \sim 10^{-2}$. 
Due to this small coupling, its production cross section in lepton as well as hadron colliders is very small. 
Further discussion of the phenomenology of this new Higgs boson is beyond the scope of this paper.  % cannot be produced in hadron colliders and is therefore fundamentally different from the SM Higgs field in its properties.
In fact, in a recent paper~\cite{nobu2}, we have argued that for some parameter ranges of the theory, the $\sigma$ particle can be a decaying dark matter of the universe.

As far as other phenomenology of the model is concerned, we note that for $g^2_{BL}/M^2_{Z_{BL}}\lesssim10^{-6}$ GeV$^{-2}$, the neutral current and other low energy constraints are automatically satisfied~(see Table 8.13 of reference \cite{LEP_rev}). This limit broadly satisfies all the LEP constraints for $VV$ type current couplings. It also implies that $g_{BL}\lesssim 10^{-3} \, M_{Z_{BL}}[{\rm GeV}]$ is allowed by low energy observations and we seek other constraints in this domain when a dark matter is included in the theory.
There are also ATLAS upper bounds on $g_{BL}$ as a function of $M_{Z_{BL}}$ but this bound for low mass $Z_{BL}$ 
is in the range of $g_{BL}\leq 2\times 10^{-3}$ or so \cite{ATLbound}  for $M_{Z_{BL}}$ about a GeV and it becomes weaker as we go to higher masses. See also the review~\cite{Lang}.

We also note that our model is different from other $U(1)'$ models since in our case the $Z_{BL}$ coupling with quarks and leptons 
is specified by the $B-L$ charges of the fermions. 
One the other hand, the DM field has an arbitrary $B-L$ charge $Q$ 
and we investigate the phenomenological viability of our model for a wide range of  the parameter space
  from $|Q| \ll 1$ to $|Q| \gg 1$.

%%%%%%%%%%%%%%%%%%%%%%%%%%%%%
\section{Case (i): Thermal dark matter constraints} 
\label{sec:3}
%%%%%%%%%%%%%%%%%%%%%%%%%%%%%
\subsection{Dark matter relic density}
We first consider the case where the parameter range of the model is such that $\zeta$ is a thermal dark matter. 
We will find these parameter ranges and their possible implications below. 
This is  the case where both $g_{BL}$ and $g_\zeta$ have such values that $Z_{BL}$, $\zeta$ 
  and SM particles were all in thermal equilibrium in the early universe, 
  followed by the  dark matter decoupling which leads to the DM relic density.
  
We first note that the dark matter interacts with the SM particle only via the $B-L$ gauge interactions. 
The Higgs boson field that breaks $B-L$ does not couple to the dark matter particle due to their $B-L$ charge mismatch 
and therefore does not contribute to the thermal equilibrium consideration between $\zeta$ and SM particles.

For the Dirac DM particle $\zeta$ to be a thermal dark matter, whose relic abundance is determined by thermal freeze-out, 
  it must be in thermal equilibrium with the SM particles as well as the $Z_{BL}$ in the very early universe. 
As the temperature of the universe drops below the $m_\zeta$, 
  the Boltzmann suppression makes the $\zeta$ particle density low and it goes out of equilibrium. 
After thermal freeze-out occurs, the DM freely expands till the current epoch and forms the dark matter of the universe.  Its current abundance is determined by the values of $g_{BL}$, $g_{\zeta}$ and $m_\zeta$. 
%In this case, there are two constraints on the coupling parameters of the model ($g_{BL}$ and $g_\zeta$):  
%  one, which determines the freeze-out and a second which determines the DM relic density. 
%We now discuss how those constraints arise and what they are.

Typically in a thermal freeze-out situation, the fact that at one point the $\zeta$  particle was in equilibrium implies constraints on the parameters $g_{\zeta}$. We have to consider different processes that can keep $\zeta$ particles in equilibrium with the SM particles.  The first one is via direct process  ${\zeta \bar{\zeta} \to f \bar{f}}$ mediated by $Z_{BL}$, which leads to
\begin{eqnarray}
n_\zeta(T) \langle \sigma v \rangle_{\zeta \bar{\zeta} \to f \bar{f}} \geq H = 
 \sqrt{\frac{\pi^2}{90} g_*} \, \frac{T^2}{M_P}, 
\label{th_zeta}
\end{eqnarray} 
where $ n_{\zeta}(T)=\frac{3 \zeta(3)}{\pi^2} T^3$ is the DM number density for $T \gtrsim m_\zeta$, 
$g_*$ is the effective  number of degrees of freedom for SM particles in thermal equilibrium (we set $g_* = 106.75$ in the following analysis),
and $M_P=2.43 \times 10^{18}$ GeV is the reduced Planck mass. 
Since we are interested in a low mass $Z_{BL}$ boson, we obtain 
  $ \langle \sigma v \rangle \simeq \frac{ g^2_{BL} g^2_\zeta }{4 \pi T^2}$ for the $\zeta\bar{\zeta} \to f\bar{f}$ process, 
  independently of the $Z_{BL}$ mass. 
Requiring the thermal equilibrium condition to be satisfied at $T \simeq m_\zeta$, 
we obtain the following constraint on the gauge coupling parameters: 
\begin{eqnarray}
g^2_{BL}g^2_\zeta \geq 43 \, \frac{m_\zeta}{M_P}. 
\label{eq1}
\end{eqnarray}
As we will see in the next subsection, the above thermal equilibrium condition is not consistent
with the direct DM detection constraints which are very severe for low $M_{Z_{BL}}$.

The second possibility for $\zeta$ to be in equilibrium with the SM particles is via a two step process:  in the 
first step $Z_{BL}$ comes to equilibrium with SM fermions via the process $f \bar{f} \to Z_{BL}\gamma$ and then
$\zeta$ goes into equilibrium with $Z_{BL} $ and hence with the SM fermions via the process 
$Z_{BL}Z_{BL} \to \zeta \zeta$. 
The thermal equilibrium condition for the first process is 
\begin{eqnarray}
n_{Z_{BL}}(T) \langle \sigma v \rangle_{f \bar{f} \to Z_{BL} \gamma} \geq H = 
 \sqrt{\frac{\pi^2}{90} g_*} \, \frac{T^2}{M_P}, 
\end{eqnarray} 
where $ n_{Z_{BL}}(T)=\frac{2 \zeta(3)}{\pi^2} T^3$ is the number density of $Z_{BL}$,
and $\langle \sigma v \rangle_{f \bar{f} \to Z_{BL} \gamma} \simeq  \frac{g_{BL}^2 \alpha_e}{T^2}$
with the fine-structure constant of $\alpha_e=1/128$. 
We require that this condition is satisfied at $T = m_\zeta$ (at latest) and obtain 
\begin{eqnarray} 
 g_{BL}\geq 2.7 \times 10^{-8} \sqrt{m_\zeta[{\rm GeV}]}  .
\label{lbd1}
\end{eqnarray}
The second process depends only on $g_\zeta$ and the equilibrium condition gives a lower bound on 
$g_\zeta \geq 9.2 \times 10^{-5}  \,  \left(  m_\zeta [{\rm GeV}] \right)^{1/4}$ 
by using $\langle \sigma v \rangle_{Z_{BL} Z_{BL} \to  \zeta \bar{\zeta}} \simeq \frac{g_\zeta^4}{16 \pi T^2}$ in Eq.~(\ref{th_zeta}). 
Clearly if we want to get the DM relic density right, we need a larger $g_\zeta$ and therefore it is in our acceptable range for the DM relic density, $\zeta$ is in thermal equilibrium with $Z_{BL}$.  Note that  the processes $Z_{BL}\to ff$ and $ff\to Z_{BL}$ which apparently are not suppressed by electromagnetic coupling, are expected to be phase space suppressed instead; so we do not consider them here.

Next, we  discuss the DM relic density constraints on the model. 
To evaluate the DM relic density, we solve the Boltzmann equation given by 
\begin{eqnarray} 
  \frac{dY}{dx}
  = - \frac{\langle \sigma v \rangle}{x^2}\frac{s (m_\zeta)}{H(m_\zeta)} \left( Y^2-Y_{EQ}^2 \right), 
\label{Boltzmann}
\end{eqnarray}  
where $x=m_\zeta/T$ is  the inverse ``temperature'' normalized by the DM mass $m_\zeta$, 
  $\langle \sigma v \rangle$ is a thermally averaged DM annihilation cross section ($\sigma$) times relative velocity ($v$), 
  $H(m_{\zeta})$ is the Hubble parameter at $T=m_{\zeta}$, 
  $s(m_{\zeta})$ is the entropy density of the thermal plasma at $T=m_{\zeta}$,  
  $Y$ is the yield of the DM particle which is defined as a ratio of the DM number density to the entropy density, 
  and $Y_{EQ}$ is the yield of the DM in thermal equilibrium. 
Explicit forms for the quantities in the Boltzmann equation are as follows:   
\begin{eqnarray} 
 H(m_{\zeta}) &=&  \sqrt{\frac{\pi^2}{90} g_*} \frac{m_{\zeta}^2}{M_P}, \nonumber \\
 s(m_{\zeta}) &=& \frac{2  \pi^2}{45} g_* m_{\zeta}^3,  \nonumber \\
 Y_{EQ}(x) &=&  \frac{g_{DM}}{2 \pi^2} \frac{x^2 m_{\zeta}^3}{s(m_{\zeta})} K_2(x),   
\end{eqnarray}
where $K_2 (x) $ is the modified Bessel function of the second kind, and 
   $g_{DM}=4$ is the number of degrees of freedom for the Dirac fermion DM particle $\zeta$. 
   
The thermal average of the DM annihilation cross section is given by the following integral expression: 
\begin{eqnarray} 
\langle \sigma v \rangle =  \frac{g_{DM}^2}{64 \pi^4}
  \left(\frac{m_{\zeta}}{x}\right) \frac{1}{n_{EQ}^{2}}
  \int_{4 m_{\zeta}^2}^\infty  ds \; \left( \sigma v  \right) s \sqrt{s-4 m_\zeta^2} \, K_1 \left(\frac{x \sqrt{s}}{m_{\zeta}}\right),
\label{ThAvgSigma}
\end{eqnarray}
where $n_{EQ}=s(m_{\zeta}) Y_{EQ}/x^3$ is the DM number density, and 
  $K_1$ is the modified Bessel function of the first kind. 
The DM annihilation occurs via the process 
$\zeta \bar{\zeta} \to Z_{BL} Z_{BL}$ for $m_\zeta > M_{Z_{BL}}$. 
%%%%%%%%%%%%%%%%%%%
In our considerations above, we have ignored the inverse decay process $\zeta\bar{\zeta}\to Z_{BL}$, since it
is a small contribution at high temperatures,  suppressed by a very small volume of the phase space. 
We have also not taken into account the Sommerfeld enhancement. 
Typically, Sommerfeld enhancement is significant if the DM speed is very low and bound states 
  of $\zeta$-$\bar{\zeta}$ are formed with a large $g_{\zeta}$ value. 
In our freeze-out scenario, the annihilation cross section in the early universe uses the speed $v\sim 0.05$ or so and the coupling is not so large (see Eq.~(\ref{DM-approx})). 
Similarly, the condition for DM bound state formation is not satisfied. 
We estimate the Sommerfeld enhancement factor to be therefore small at the freeze-out epoch. 
However, at the recombination and the current epoch, Sommerfeld effect is significant due to very low velocities of DM particles 
and leads to important constraints on the parameters for the freeze-out case (see below).
%%%%%%%%%%%%%%%%%%%%%%%%%%%%%
By solving the Boltzmann equation of Eq.~(\ref{Boltzmann}) with the initial condition 
  $Y(x)=Y_{EQ}(x)$ for $x \ll1$, we evaluate the DM yield at present, $Y(x \to \infty)$. 
The relic abundance of the DM in the present universe is then given by
\begin{eqnarray}
  \Omega_{DM} \,h^2 =\frac{m_{\zeta} s_0 Y(\infty)} {\rho_c/h^2}, 
  \label{presentDM}
\end{eqnarray} 
where $s_0 = 2890$ cm$^{-3}$ is the entropy density of the present Universe, 
   and $\rho_c/h^2 =1.05 \times 10^{-5}$ GeV/cm$^3$ is the critical density.  
For the thermal DM scenario, the asymptotic solution of the Boltzmann equation ($Y(\infty)$) is known, 
  and with a good accuracy, the thermal DM relic density is expressed to be~\cite{review1, review2}
\begin{eqnarray}
\Omega_{DM} \, h^{2}
%\frac{m_{DM} \, s_{0} \, Y (\infty)}{\rho_{c}/h^2}
%\; ,
\simeq \frac{ 2.13 \times 10^{8} \, x_{f}}{\sqrt{g_{\ast}} \, M_{P} \, \langle \sigma v \rangle} \; ,
 \label{Omega}
\end{eqnarray}
where $M_P$ and $\langle \sigma v \rangle$ are evaluated in units of GeV, 
the freeze-out temperature of the DM particle is approximately evaluated as
 $x_{f}=m_{\zeta}/T_{f}\simeq \ln(x)-0.5\ln(\ln(x))$
 with $x \simeq 0.19 \sqrt{g_{DM}/g_{*}} M_{P} \, m_{\zeta} \, \langle  \sigma v \rangle$.
Since the annihilation process occurs via $s$-wave, we can approximate $\langle \sigma v \rangle$ as $\sigma v $ 
  in the non-relativistic limit. 
Here in our analysis, we employ Eqs.~(\ref{Eq:App5}) and (\ref{Eq:App2})
given in Appendix for the annihilation processes $\zeta \bar{\zeta}  \to Z_{BL} Z_{BL}$
and $\zeta \bar{\zeta}  \to f \bar{f}$, respectively. 
As we will discuss in the following subsection, the direct DM detection constraints 
are very severe and we find that they require $g_{BL} \ll g_\zeta$. 
Thus, the contribution from the process $\zeta \bar{\zeta}  \to f \bar{f}$ is negligibly small.

%%%%%%%%%%%%%%%%%%%%%%%%%%%%%%%%%%%%%%%%%%%%%%%
\begin{figure}[t]
  \centering
 \includegraphics[width=0.6\linewidth]{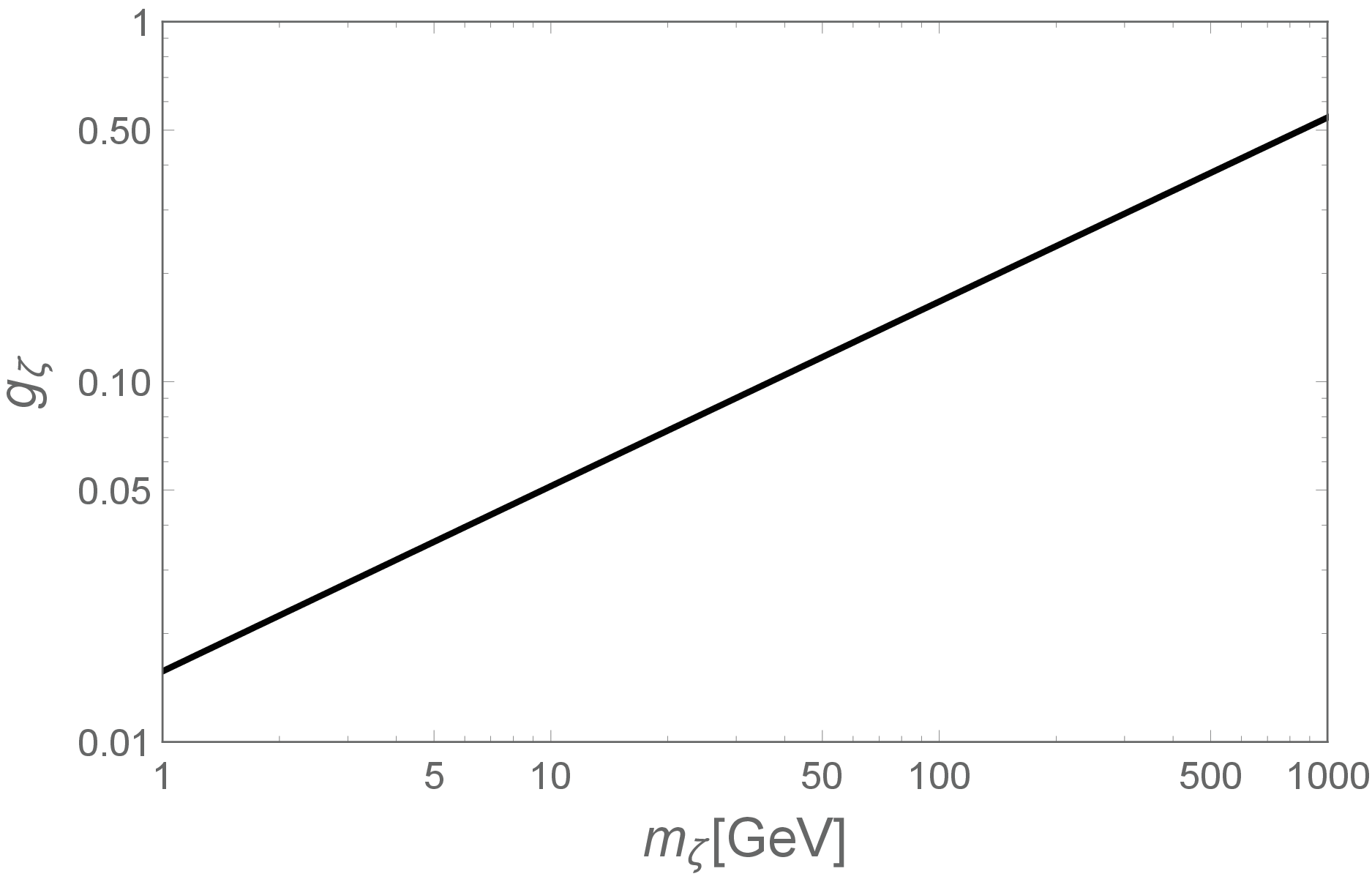}
  \caption{The relation between the DM mass and the DM coupling with $Z_{BL}$
for the case of $M^2_{Z_{BL}}  \ll m^2_\zeta$. 
The observed DM relic density is reproduced along the line. 
$g_\zeta \simeq  0.016 \times \sqrt{m_\zeta[{\rm GeV}]}$ is a good approximation formula. 
 }
 \label{DMV}
\end{figure}
%%%%%%%%%%%%%%%%%%%%%%%%%%%%%%%%%%%%%%%%%%%%%%%

\begin{figure}[t]
  \centering
  \includegraphics[width=0.6\linewidth]{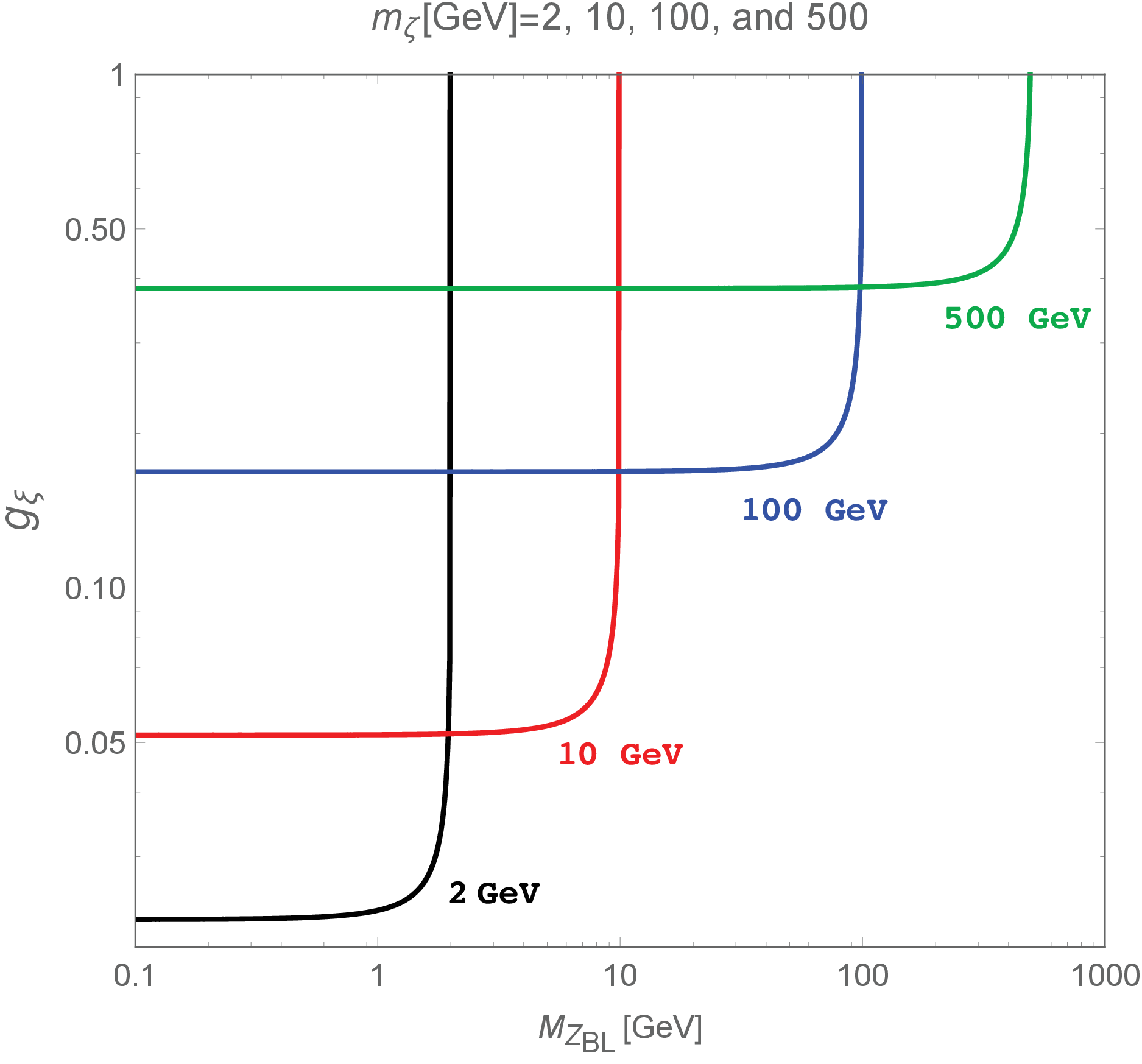}
  \caption{The relation between the $Z_{BL}$ mass and the DM coupling with $Z_{BL}$
for fixed DM masses of 2 GeV (black), 10 GeV (red), 100 GeV (blue), and 500 GeV (green). 
The observed DM relic density is reproduced along each line. 
For $M^2_{Z_{BL}} \ll m^2_{\zeta}$, the coupling is almost constant for a fixed DM mass. 
The coupling rises sharply when the $Z_{BL}$ mass becomes very close to the DM mass
   because of  phase space effect. 
}
\label{DMV2}   
  \end{figure}

In order to reproduce the observed DM relic density at the present epoch, $\Omega_{DM}h^2 = 0.12$ \cite{Planck2019}, 
  we obtain a relation between the DM mass and the DM coupling with $Z_{BL}$ 
  for $M^2_{Z_{BL}}  \ll m^2_\zeta$, which is shown by the line in Fig.~\ref{DMV}. 
The observed DM relic density is reproduced along the line, which we find to be well approximated by
\begin{eqnarray} 
  g_\zeta \simeq  0.016 \times \sqrt{m_\zeta [{\rm GeV}]}. 
\label{DM-approx}
\end{eqnarray}
As we expected, the thermal equilibrium condition for the process $Z_{BL} Z_{BL} \leftrightarrow \zeta \bar{\zeta}$ 
  we have found before (see after Eq.~(\ref{lbd1})) is always satisfied for $m_\zeta > 1$ GeV.  
In Fig.~\ref{DMV2}, we show the relation between the $Z_{BL}$ mass and the DM coupling with $Z_{BL}$
   for fixed DM masses of 2 GeV (black), 10 GeV (red), 100 GeV (blue), and 500 GeV (green). 
The observed DM relic density is reproduced along each line. 
We can see that the coupling is almost constant for a fixed DM mass for $M^2_{Z_{BL}} \ll m^2_{\zeta}$ 
  and is well-approximated by Eq.~(\ref{DM-approx}). 
The coupling is sharply rising when the $Z_{BL}$ mass becomes very close to the DM mass
   because of the phase space/kinematic effect.

%After freeze-out, the dark matter fermion density relative to the entropy density is given by 
%\begin{eqnarray}
%\frac{n(T_F)}{s}\sim \frac{g^{*1/2}}{<\sigma v >T_F M_P}
%\end{eqnarray}
%and the DM freely floats and dilutes as the universe expands while keeping the $\frac{n_\zeta(T_F}{s}$ constant (where $s$ is the entropy density of the universe). Using the current entropy of the universe, we find that to fit the observed dark matter density of $\Omega_{DM}h^2\sim 0.1$, we get
%\begin{eqnarray}
%\Omega_{DM}\simeq 0.3\simeq\left( \frac{g^{*1/2}m_\zeta}{<\sigma v > T_F M_P}\right)\left(\frac{3000/cm^3}{1.6\times 10^{-5}~GeV/cm^3}\right)
%\label{eq2}
%\end{eqnarray}
%Using the above expression for $<\sigma v>\simeq \frac{{g^4}_\zeta}{4\pi m^2_\zeta}$ and the fact that freeze out temperature $T_F\sim \frac{m_\zeta}{20}$ in Eq.\ref{eq2} , we get $g_\zeta\simeq 0.04\sqrt{m_\zeta/{\rm GeV}}$ to get the correct dark matter density right now (see Fig.\ref{DMV}.

%%%%%%%%%%%%%%%%%%%%%%%%%%%%%%%%%%%%%%%%%%%%%%%%
%%%%%%%%%%%%%%%%%%%%%%%%%%%%%%%%%%%%%%%%%%%%%%%%

%Using this value for $g_\zeta$ in Eq.\ref{eq1}, we find that 
%\begin{eqnarray}
%g_{BL}\geq  1.6\times 10^{-7}\sqrt{m_\zeta({\rm GeV})}
%\label{lbd}
%\end{eqnarray}
%for $Z_{BL}$ and $\zeta$ to be in thermal equilibrium with the SM particles. 
%%%%%%%%%%%%
%There is, however, another possibility for $\zeta$ to be in equilibrium with the SM particles i.e. via a two step process: first $\zeta\zeta\to Z_{BL}Z_{BL}$ comes to  equilibrium and then $Z_{BL}$ goes into equilibrium with SM fermions via the process $Z_{BL}\gamma\to f\bar{f}$. The first process depends only on $g_\zeta$ and equilibrium condition for the first process gives a lower bound on $g_\zeta \geq 10^{-4}$. Clearly if we want to get the DM relic density right, we need a larger $g_\zeta$ and therefore in our acceptable range for the DM relic density, $\zeta$ is in thermal equilibrium with $Z_{BL}$. Now for the process $Z_{BL}\gamma\to f\bar{f}$ to be in thermal equilibrium, the condition can be easily derived to be 
%\begin{eqnarray}
%g_{BL}\geq 10^{-8} \sqrt{m_\zeta({\rm GeV})}.
%\label{lbd1}
%\end{eqnarray}

%%%%%%%%%%%%%%%%%%%%%%
\subsection{Direct detection  constraints}
\label{sec:3.2}
%%%%%%%%%%%%%%%%%%%%%%
%%%%%%%%%%%%%%%%%%%%%%%%%%%%%%%%%%%%%%%%%%%%%%%%
\begin{figure}[t]
  \centering
  \includegraphics[width=0.6\linewidth]{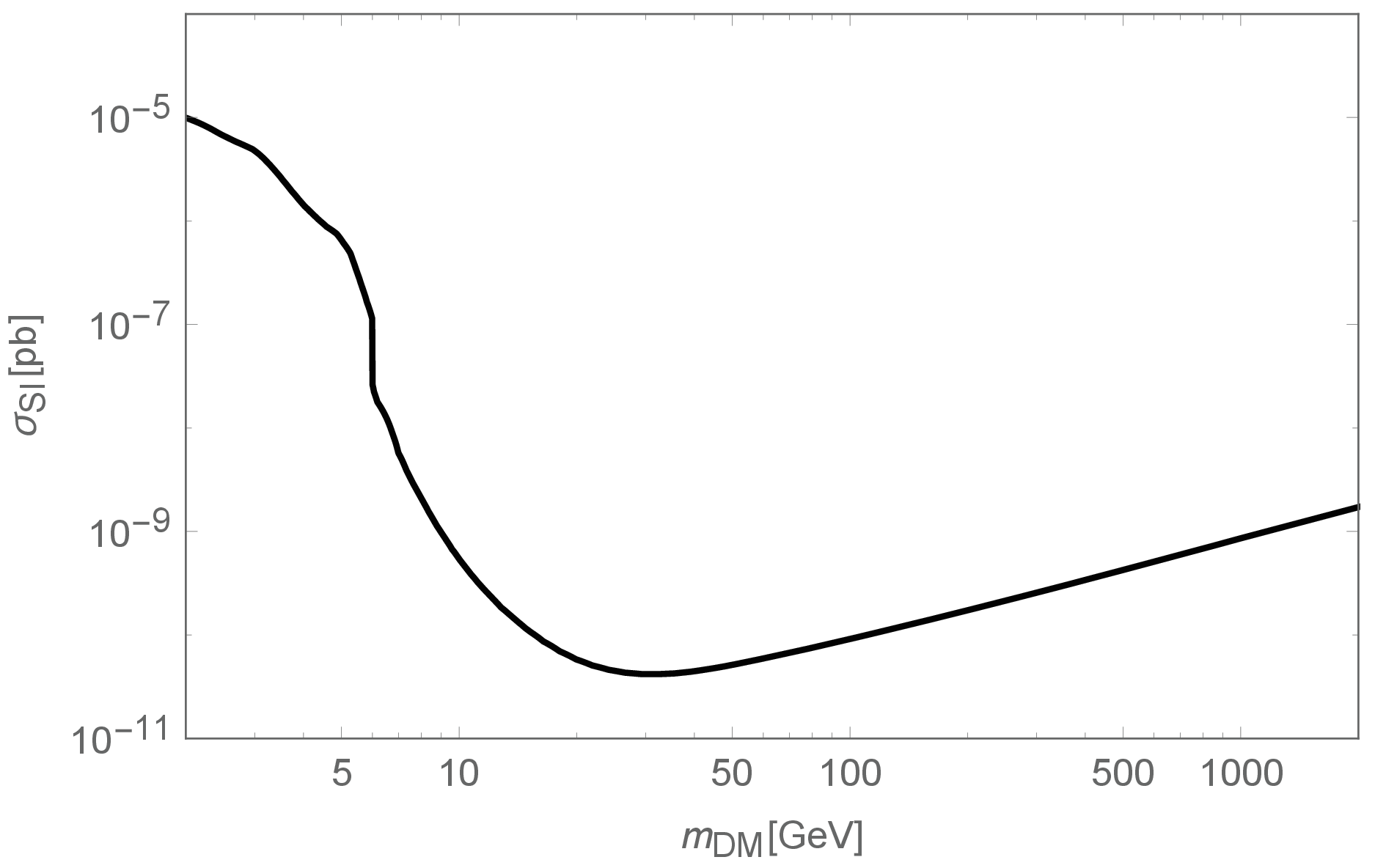}
  \caption{
The current experimental upper bound on the spin-independent cross section as a function of the DM mass.}
\label{SI-data}   
  \end{figure}
%%%%%%%%%%%%%%%%%%%%%%%%%%%%%%%%%%%%%%%%%%%%%%%%

Let us now turn to the direct detection constraints. 
In Fig.~\ref{SI-data}, we show the current upper bound on the spin-independent cross section ($\sigma_{SI}$) 
  for the elastic scattering of the DM particle with a nucleon for the DM mass of $m_{DM} \geq 2$ GeV. 
For the DM mass $m_{DM} \geq 6$ GeV, the most stringent upper bound is obtained by XENON1T experiment \cite{Xenon1T-2018}
  while for $2 \, {\rm GeV} \leq m_{DM} \leq 6$ GeV, the upper bound is obtained by a combination of 
  DarkSide-50 \cite{DarkSide-50},  LUX \cite{LUX-2019} and PandaX-II \cite{PandaX-II}. 
As is well known the constraints are most severe for a DM mass around 30 GeV and become weaker on either side of this mass.

\begin{figure}[tbh]
  \centering
   \includegraphics[width=0.6\linewidth]{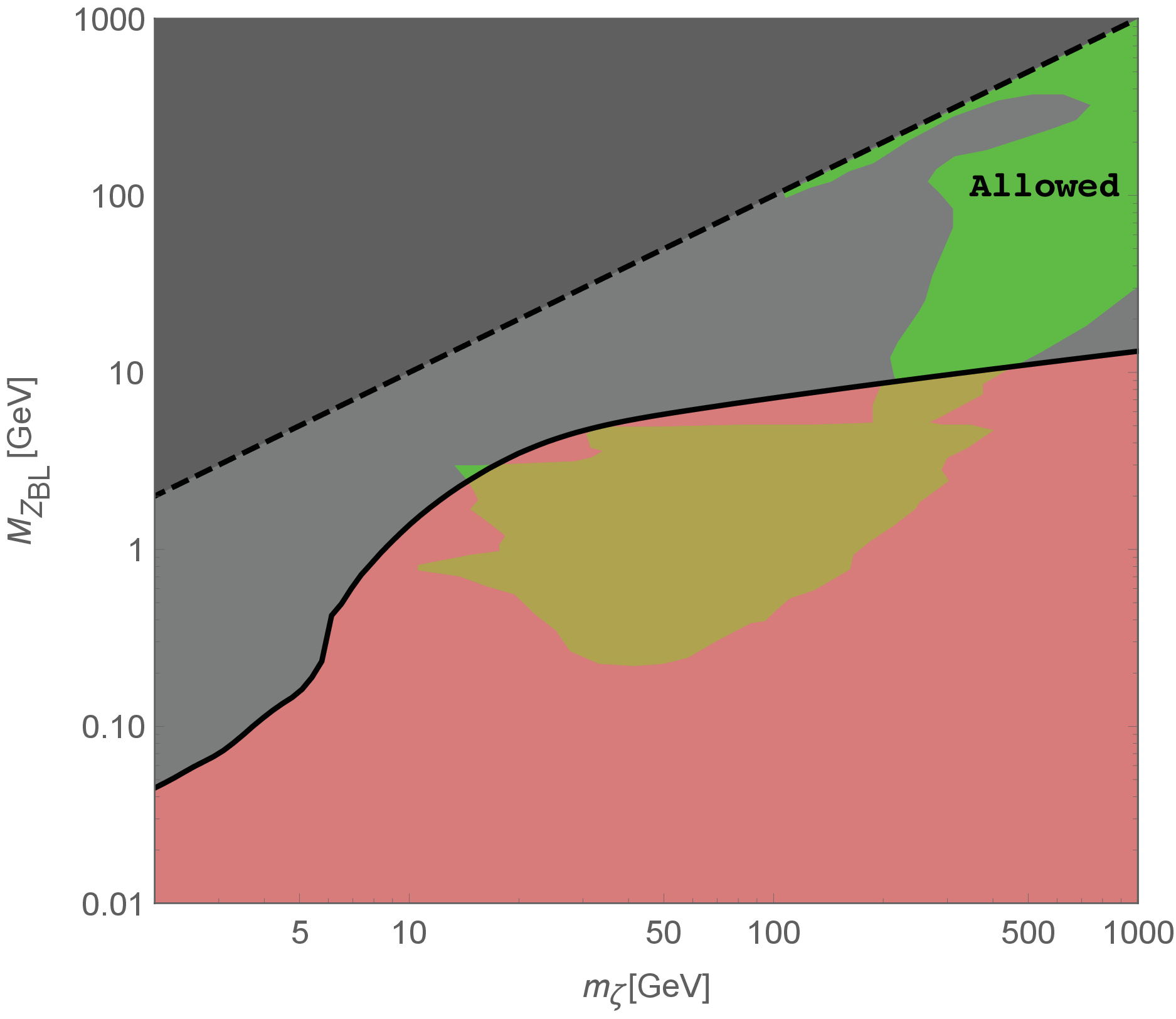}
  \caption{
The parameter regions in $(m_\zeta, M_{Z_{BL}})$-plane that satisfy the conditions 
    from the spin-independent cross section bounds and the indirect detection constraints obtained in Ref.~\cite{cirelli}, 
    as well as the condition for the thermal equilibrium between $Z_{BL}$ and the SM particles.
The region below the solid black line is disallowed by the spin-independent cross section constraints
    and the thermal equilibrium between $Z_{BL}$ and the SM particles. 
The region above the dashed line, which corresponds to $m_\zeta \leq m_{Z_{BL}}$,  is  not considered in the paper. 
The gray region is ruled out by indirect constraints from the CMB data and the AMS-02 results~\cite{cirelli}. 
The yellowish looking region is the extension of the green region and only the tip of it sticks out in the middle of the figure. 
The allowed region for the freeze-out case then turns out to be the green region i.e. 
   $m_\zeta \gtrsim 100$ GeV and $M_{Z_{BL}} \gtrsim 10$ GeV. 
} 
  \label{ZBL_min}
\end{figure}

In our model, the elastic scattering of the DM particle with a nucleon $\zeta N \to \zeta N$
  occurs via the exchange of $Z_{BL}$ boson.
The  cross section for the process is given by \cite{farinaldo}
\begin{eqnarray}
\sigma_{SI}=\frac{1}{\pi} \, g_{\zeta}^{2} \, g_{BL}^{2} \, \frac{\mu_{\zeta N}^{\, 2}}{M_{Z_{BL}}^{\, 4}} ,
\label{SI}
\end{eqnarray}
where $\mu_{\zeta N}=m_{\zeta}m_{N}/(m_{\zeta}+m_{N})$ is the reduced mass for the DM-nucleon system 
with $m_{N}=0.983$ GeV being the nucleon mass. 
Note that this cross section formula is valid for $M_{Z_{BL}}^2 \gtrsim M_T E_R$, 
  where $M_T$ is a target nuclei mass, and $E_R$ is a typical recoil energy.  
For XENON1T experiment, $M_T \sim 100$ GeV and $E_R \sim 10$ keV, 
  so that we can apply Eq.~(\ref{SI}) for $M_{Z_{BL}} \gtrsim 50$ MeV.   
As $M_{Z_{BL}}$ decreases from $M_{Z_{BL}} = 50$ MeV, 
   the $Z_{BL}$ exchange process becomes long-range 
   and $\sigma_{SI}$ quickly approaches a constant value as shown
   in Refs.~\cite{DelNobile:2015uua, DelNobile:2015bqo, Panci:2014gga, Li:2014vza}.  
For $M_{Z_{BL}} < 50$ MeV, we approximate the constant cross section by Eq.~(\ref{SI})
   with $M_{Z_{BL}} = 50$ MeV fixed. 
For a given $m_\zeta$, say, one GeV, which satisfies all the above constraints, 
we see that as $M_{Z_{BL}}$ goes down, the cross section rises in Eq.~(\ref{SI}). 
Since $g_{BL}$ has a lower bound from Eq.~(\ref{lbd1}) and $g_\zeta$ values are already fixed, 
   this implies a lower bound on $M_{Z_{BL}}$ depending on the $\zeta$ mass 
   along the upper bound on $\sigma_{SI}$ in Fig.~\ref{SI-data}. 
This lower bound is shown as the black solid line in Fig.~\ref{ZBL_min}.  
For example, for $ m_{\zeta}=2$ GeV, we find the minimum $Z_{BL}$ mass to be $\simeq 50$ MeV. 

%%%%%%%%%%%%%%%%%%%%%%%%%%%
\subsection{Indirect detection constraints}
%%%%%%%%%%%%%%%%%%%%%%%%%%%
In our model, the dark matter annihilation  to $Z_{BL}Z_{BL}$ at late time can undergo Sommerfeld enhancement due to the low velocity of  DM fermion.  
The $Z_{BL}$s can subsequently decay to SM fermions, which can lead to signals in indirect DM searches 
  such as the CMB measurement and AMS-02 anti-proton searches.  
These constraints have been analyzed in Refs.~\cite{walia, cirelli}  and they lead to very tight constraints on DM mass 
  in the range of 1 GeV to 100 GeV. Even though the Ref.~\cite{cirelli} considers a dark photon portal, it is very similar to our $B-L$ portal  and therefore we can apply their constraints to our case.
In Fig.~4, we have combined the direct detection constraint with the indirect detection constraints obtained in Ref.~\cite{cirelli}.  
The green region is allowed by all the constraints and  this pretty much rules out the low mass (thermal) DM scenario 
  for $m_{Z_{BL}} \lesssim 10$ GeV.

%%%%%%%%%%%%%%%%%%%%%%%%%
%%%%%%%%%%%%%%%%%%%%%%%%%%%%%%%%

%%%%%%%%%%%%%%%%%%%%%%%%%%

%%%%%%%%%%%%%%%%%%%%%%%%%%%%%
\section{Case (ii): Freeze-in dark matter scenario}  
\label{sec:4}
%%%%%%%%%%%%%%%%%%%%%%%%%%%%%
In this case, we require  the dark matter fermion $\zeta$  not to be in equilibrium with either the SM particles or the $Z_{BL}$. There are then several constraints on the couplings $g_{BL}$ and $g_\zeta$ that emerge in this case if $\zeta$ has to play the role of dark matter. We discuss them below.
\subsection{Dark matter relic density}
%%%%%%%%%%%%%%%%%%%%

This case arises when the gauge couplings $g_{BL}$ and $g_\zeta$ have much smaller values
  than the freeze-out case so that the dark matter particle was never in equilibrium with the thermal plasma of the SM particles. 
In this section, we assume that the $\zeta$ particle had zero initial abundance at the reheating after inflation. 
Productions of $\zeta$ particles from inflaton decay will be briefly discussed in Sec.~\ref{sec:5}. 
There are then two possible cases: 

(A)  the $Z_{BL}$ was in thermal equilibrium with SM particles. 
This corresponds to the case where $g_{BL}\geq 2.7 \times 10^{-8} \sqrt{m_\zeta[{\rm GeV}]}$, and  

(B) the $Z_{BL}$ was not in thermal equilibrium with SM particles i.e.~$g_{BL}< 2.7 \times 10^{-8} \sqrt{m_\zeta[{\rm GeV}]}$. 

\noindent For case (A), we find that
the most conservative conditions for the reaction $\zeta\bar\zeta \leftrightarrow f\bar{f}$ 
to be out of equilibrium till the BBN epoch is: 
\begin{eqnarray}
g_{BL} \, g_\zeta \leq 10^{-10}
\end{eqnarray}
This follows for $M_{Z_{BL}} \leq 1$ GeV  and requiring that the above reaction falls out of equilibrium 
above $T=1$ GeV epoch of the universe. For higher $Z_{BL}$ masses, the condition is even weaker.
%and $\zeta$ being out of equilibrium with $Z_{BL}$ is given by $g_\zeta \leq 10^{-4.5}$.
Similarly, for DM mass is in the low GeV range,  there is Boltzmann suppression in its number density 
and the bound becomes weaker as well. 
Similarly, for the process $\zeta\bar\zeta \leftrightarrow Z_{BL}Z_{BL}$, the corresponding condition is
\begin{eqnarray}
 g_\zeta \leq 9.2 \times 10^{-5}  \,  \left(  m_\zeta [{\rm GeV}] \right)^{1/4},  
\end{eqnarray}

Next, we proceed to evaluate the DM relic abundance by numerically solving the Boltzmann equation in Eq.~(\ref{Boltzmann}). 
Note that even for the freeze-in case the Boltzmann equation is of the same form as in the thermal dark matter case. 
This is because the term proportional to $Y_{EQ}^2$ in the right-hand side of Eq.~(\ref{Boltzmann}) 
corresponds to the DM particle productions from the SM thermal plasma. 
The difference from the thermal dark matter case is that we set the boundary condition for the freeze-in case to be $Y(x_{RH})=0$, 
where $x_{RH}=m_{\zeta}/T_{RH} \ll1 $ is related to the reheat temperature ($T_{RH}$) after inflation. 
The relic abundance of the DM in the present universe is given in Eq.~(\ref{presentDM}). 

In evaluating the thermal average of the DM annihilation cross section in Eq.~(\ref{ThAvgSigma}), 
  we consider two processes for the DM particle creation, 
  $ f \bar{f} \to \zeta \bar{\zeta} $ mediated by $Z_{BL}$ and $ Z_{BL} Z_{BL} \to \zeta \bar{\zeta}$. 
Note that the second process is active only for case (A) 
  (except for a special case, sequential freeze-in, that we discuss below).  
The corresponding cross sections are given by those of the DM annihilation processes. 
In Appendix, we list the exact cross section formulas for the processes. 
Using them for Eq.~(\ref{ThAvgSigma}), we evaluate the thermal average of the cross section
  and then numerically solve the Boltzmann equation of Eq.~(\ref{Boltzmann}) 
  with the boundary condition of $Y(x_{RH})=0$. 
In the freeze-in mechanism, the DM particles are created mostly in the relativistic regime, $T \gg m_\zeta$,
  where the annihilation cross sections are approximately given by (see Eqs.~(\ref{Eq:App3}) and (\ref{Eq:App6}) in Appendix)
\begin{eqnarray} 
&& \sigma (\bar{\zeta} \zeta \to f \bar{f} )\, v \simeq  \frac{37}{36 \pi s}  g_\zeta^2 g_{BL}^2,  \nonumber\\
&& \sigma (\bar{\zeta} \zeta \to Z_{BL} Z_{BL})\, v \simeq  \frac{g_\zeta^4}{4 \pi s } \left( \ln \left[\frac{s}{m_\zeta^2} \right] -1 \right), 
\label{DMSigma}
\end{eqnarray} 
where we have assumed $m_b^2 \ll m_\zeta^2 < m_t^2$ and $m_{Z_{BL}}^2 \ll m_\zeta^2$.  
Although we use the exact cross section formulas to evaluate $Y(x)$ in our analysis,   
  we find that the approximation formulas in Eq.~(\ref{DMSigma}) lead to 
  almost the same results as those obtained by the exact formulas.

\begin{figure}[t]
\begin{center}
\includegraphics[width=0.7\linewidth]{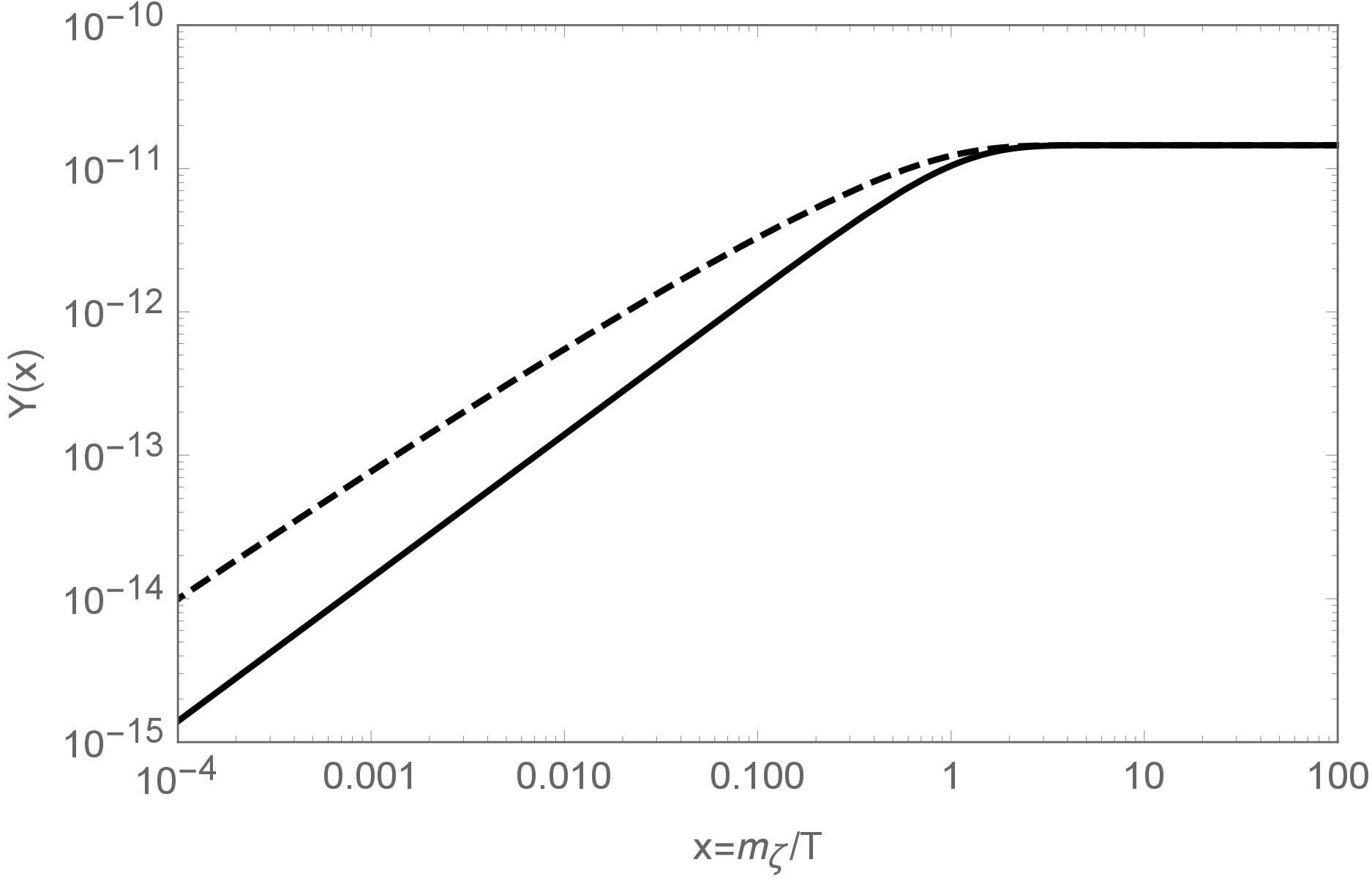}
\end{center}
\caption{
The yield of the Dirac DM particle as a function of $x=m_\zeta/T$ for $m_\zeta=30$ GeV. 
The solid line denotes the result for  $g_\zeta^2  \, g_{BL}^2  = 8.2 \times 10^{-24}$ 
   with  $g_{BL}^2 \gg g_\zeta^2$, 
   while the dashed line denotes the result for  $g_\zeta^4 = 1.2 \times 10^{-23}$ 
   with  $g_{BL}^2 \ll g_\zeta^2$. 
In both cases, $\Omega_{DM} h^2=0.12$  is reproduced. 
In this analysis, we have used the exact formulas for the annihilation cross sections 
  given in Appendix. 
}
\label{fig:Y}
\end{figure}

In Fig.~\ref{fig:Y}, fixing $m_\zeta=30$ GeV, we show the resultant $Y(x)$ for two cases: 
One is for $g_\zeta^2  g_{BL}^2 =8.2  \times 10^{-24}$ with $g_{BL} \gg g_\zeta$ (solid line), 
and the other is $g_\zeta^4 =1.2 \times 10^{-23}$ with $g_{BL} \ll g_\zeta$  (dashed line). 
For the first case, the process $\zeta \bar{\zeta}  \to f \bar{f}$ dominates, 
  while the process $\zeta \bar{\zeta} \to Z_{BL} Z_{BL}$ dominates for the second case. 
As we can see, $Y(x)$ grows from $Y(x_{RH} \ll 1)=0$ and becomes constant at $x \simeq 1$. 
Using the approximation formulas in Eq.~(\ref{DMSigma}), this behavior can be qualitatively understood as follows:  
In the first case, for $x \lesssim 1$, we have $\langle \sigma v \rangle \propto g_\zeta^2 \, g_{BL}^2 (x^2/m_\zeta^2)$, 
 and Eq.~(\ref{Boltzmann}) can be easily solved with $Y \ll Y_{EQ} \simeq$ constant and $Y(x_{RH} \ll 1)=0$.  
We find a solution to be $Y(x) \propto g_\zeta^2 g_{BL}^2 (x-x_{RH})/m_\zeta \simeq  g_\zeta^2 \, g_{B}^2 (x/m_\zeta)$. 
Since the DM particle creation from the thermal plasma should stop at $T \sim m_\zeta$ because of the kinematics, 
$Y(\infty) \sim Y(x \simeq1) \propto g_\zeta^2 g_{BL}^2 /m_\zeta$. 
Using Eq.~(\ref{presentDM}), we find that the resultant DM relic density is proportional to $g_\zeta^2 \, g_{B}^2$ 
  while independent of the DM mass.  
We have arrived at the same conclusion even for the numerical result by using the exact cross section formulas. 
We find a similar result for the second case, namely, the resultant DM relic density is proportional to $g_\zeta^4$ 
  while independent of the DM mass.  

%%%%%%%%%%%%%%%%%%%%%%%%%%%
%%%%%%%%%%%%%%%%%%%%%%%%%%%

By numerically solving the Boltzman equation,  we find that independently of $m_\zeta$, the observed DM relic density of 
$\Omega_{DM} h^2=0.12$  is reproduced in case (A) by 
\begin{eqnarray}
  g^2_\zeta \, g^2_{BL} + \frac{0.82}{1.2} \, g^4_\zeta \simeq 8.2 \times 10^{-24} 
  ~~~{\rm for}~~~ g_{BL}\geq 2.7 \times 10^{-8}\sqrt{m_\zeta[{\rm GeV}]}. 
 \label{caseA}
\end{eqnarray}  
In case (B), on the other hand, there is no $Z_{BL}$ initially, the condition is given by only the first term in the above equation, i.e. 
 \begin{eqnarray}
 g_\zeta^2 \, g_{BL}^2  \simeq 8.2 \times 10^{-24} 
 ~~~{\rm for}~~~ g_{BL} < 2.7 \times 10^{-8}\sqrt{m_\zeta[{\rm GeV}]}. 
 \label{case-b}
\end{eqnarray}  
For example, for $m_\zeta =1 $ GeV, the first equation implies that $g_\zeta \sim 10^{-6}$ or lower whereas the second case corresponds to $g_\zeta \sim 10^{-4}$ or higher.

%%%%%%%%%%%%%%%%%%%%%%%%%%%
\begin{figure}[t]
\begin{center}
\includegraphics[width=0.7\linewidth]{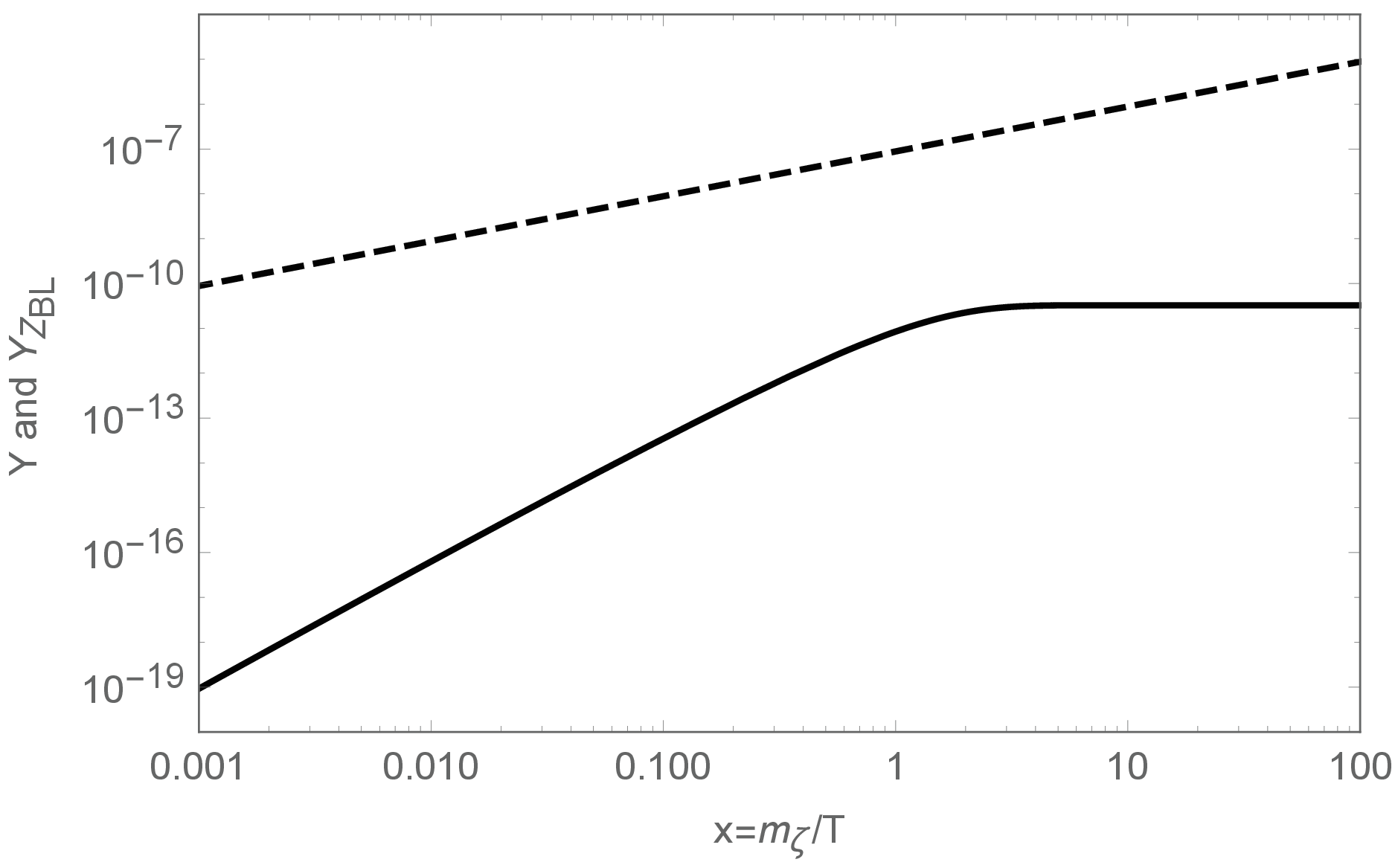}
\end{center}
\caption{
In the sequential freeze-in case, 
  the yield of the Dirac DM particle as a function of $x=m_\zeta/T$ for $m_\zeta=30$ GeV (solid line),
  along with the yield of $Z_{BL}$ (dashed line). 
Here, we have taken $g_{BL}=5.0 \times 10^{-10}$ and $g_\zeta=6.3 \times 10^{-4}$, 
  by which $\Omega_{DM} h^2=0.12$ is reproduced. 
}
\label{fig:Y2}
\end{figure}
%%%%%%%%%%%%%%%%%%%%%%%%%%%

%%%%%%%%%%%%%%%%%%%%%%%%%%%%  
\begin{figure}[t]
\begin{center}
\includegraphics[width=0.3\textwidth, ,angle=0,scale=1.55]{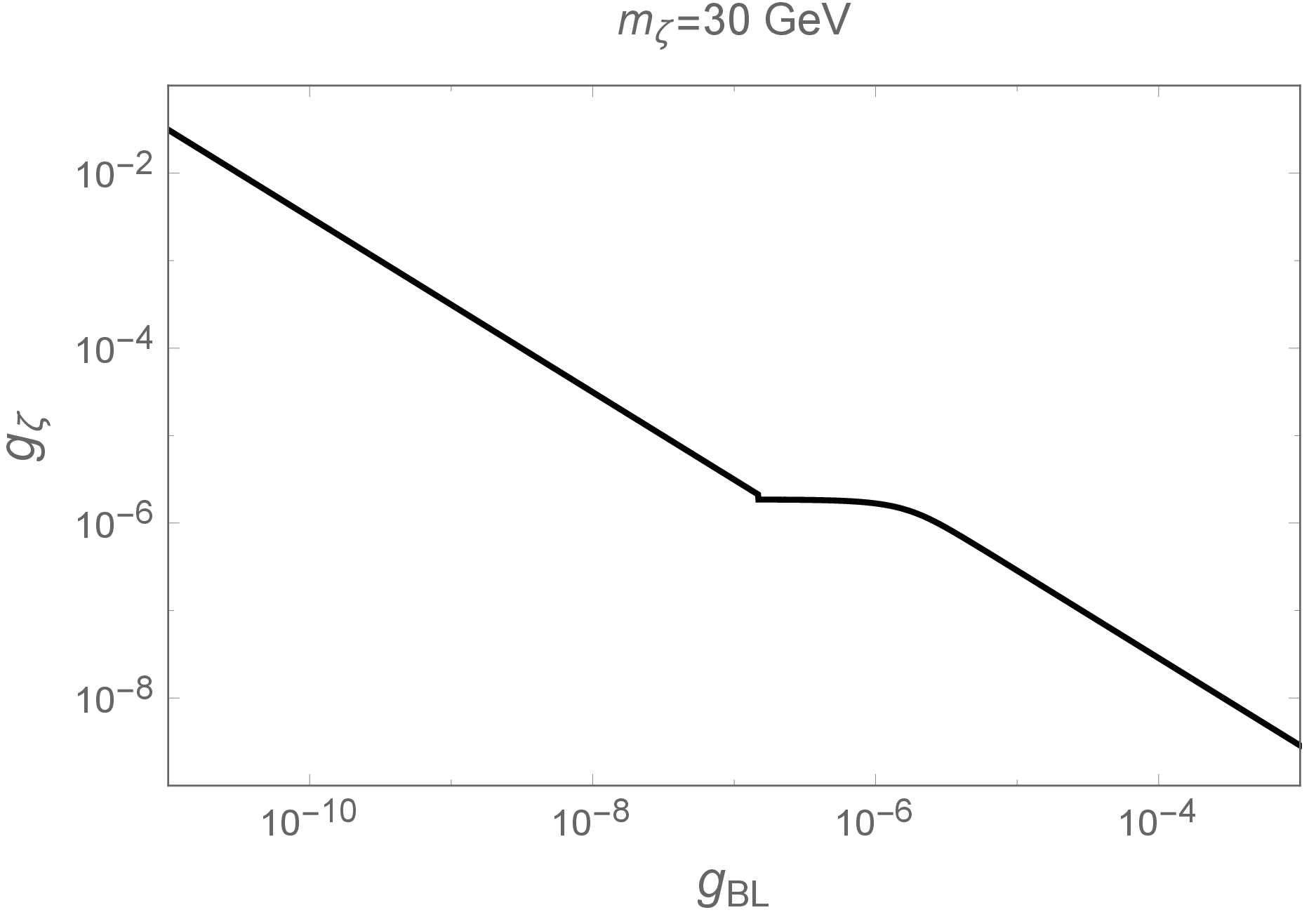} 
\includegraphics[width=0.3\textwidth, angle=0,scale=1.675]{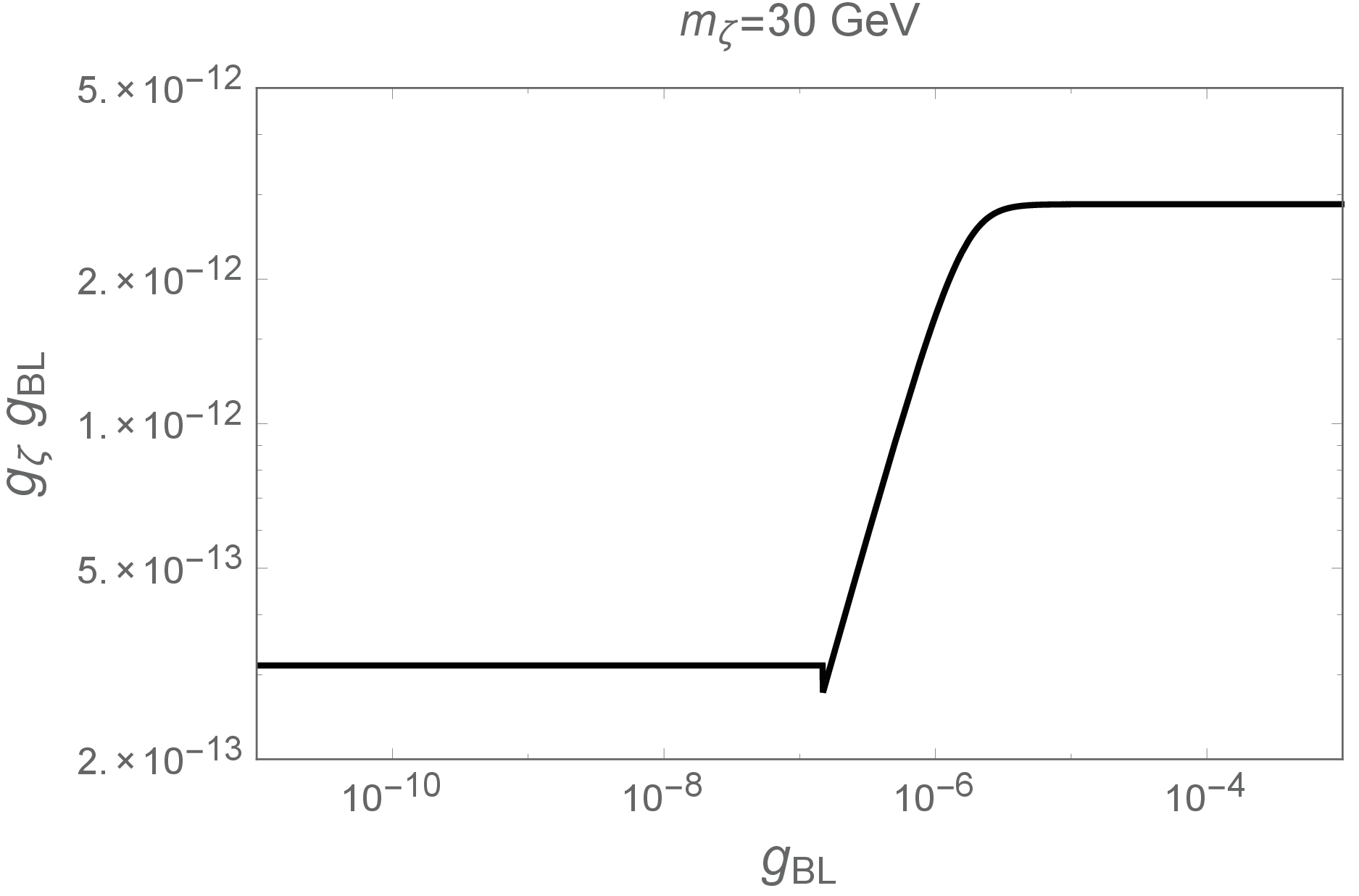}
\end{center}
\caption{The plot of $g_\zeta$ vs $g_{BL}$ (left panel) and $g_\zeta g_{BL}$ vs $g_{BL}$ (right panel) for $m_\zeta=30$ GeV. 
The observed DM relic density is reproduced along the solid lines. 
Note that since we have analyzed case (A) and case (B) separately, 
  the discontinuity appears at $g_{BL} \simeq 1.5 \times 10^{-7}$ for $m_\zeta=30$ GeV, 
  where $Z_{BL}$ goes out of/in thermal equilibrium with the SM particles. 
}
\label{Y1}
\end{figure}
%%%%%%%%%%%%%%%%%%%%%%%%%%%%%

Very recently, it has been pointed out in Ref.~\cite{Hambye:2019dwd} 
  that in case (B)  ``sequential freeze-in''  can dominantly produce
  the DM particles compared to the process of $f \bar{f} \to \zeta \bar{\zeta}$ considered above. 
If this is the case, Eq.~(\ref{case-b}) is not the right condition to reproduce $\Omega_{DM} \, h^2=0.12$. 
In the case of sequential freeze-in, the DM particles are produced in two steps. 
First, $Z_{BL}$ is produced from the thermal plasma of the SM particles, and then 
  the DM particles are produced through $ Z_{BL} Z_{BL} \to \zeta \bar{\zeta}$. 
Let us now estimate the DM relic density through the sequential freeze-in. 
The yield of $Z_{BL}$ ($Y_{Z_{BL}}$) is calculated by the Boltzmann equation, 
\begin{eqnarray} 
  \frac{dY_{Z_{BL}}}{dx}
  \simeq  \frac{\langle \sigma v \rangle_{f\bar{f} \to Z_{BL}\gamma}}{x^2}\frac{s (m_\zeta)}{H(m_\zeta)} 
   \, Y^{EQ}_{Z_{BL}} \, Y^{EQ}_\gamma, 
\label{BoltzmannBL}
\end{eqnarray}  
where $Y^{EQ}_{Z_{BL}} \simeq Y^{EQ}_\gamma \simeq 5.2 \times 10^{-3}$ 
  are the yields of $Z_{BL}$ and the photon, respectively, in the thermal equilibrium, 
  and $\langle \sigma v \rangle_{f\bar{f} \to Z_{BL}\gamma} \simeq \frac{g_{BL}^2 \alpha_e}{m_\zeta^2} x^2$. 
This Boltzmann equation is easily solved from $x_{RH} \ll 1$, and we find 
\begin{eqnarray} 
 Y_{Z_{BL}}(x) \simeq  2.9 \times 10^{-6}  \left( \frac{M_P}{m_\zeta} \right) g_{BL}^2 \, x 
\label{BoltzmannBLY}
\end{eqnarray}  
for $x \lesssim m_\zeta/M_{Z_{BL}}$. 
With this $Y_{Z_{BL}}(x)$, we calculate the DM density by solving  the Boltzmann equation, 
\begin{eqnarray} 
  \frac{dY}{dx}
&\simeq&  \frac{\langle \sigma v \rangle_{Z_{BL} Z_{BL} \to \zeta \bar{\zeta}}} {x^2}\frac{s (m_\zeta)}{H(m_\zeta)} \, Y_{Z_{BL}}^2 
\nonumber\\
&=&   
\frac{\langle \sigma v \rangle_{\zeta \bar{\zeta} \to Z_{BL} Z_{BL}}} {x^2}\frac{s (m_\zeta)}{H(m_\zeta)} \, 
Y_{EQ}^2 \, \left(\frac{Y_{Z_{BL}}}{Y_{Z_{BL}}^{EQ}}\right)^2, 
\label{BoltzmannDM}
\end{eqnarray}  
where we have used $\langle \sigma v \rangle_{Z_{BL} Z_{BL} \to \zeta \bar{\zeta}} 
(Y_{Z_{BL}}^{EQ})^2 =
\langle \sigma v \rangle_{\zeta \bar{\zeta} \to Z_{BL} Z_{BL}} \, Y_{EQ}^2$
in the second line. 
%$\langle \sigma v \rangle_{Z_{BL} Z_{BL} \to \zeta \bar{\zeta}} \simeq \frac{g_\zeta^4}{16 \pi m_\zeta^2} x^2$. 
In our analysis here, we have assumed that the sequential freeze-in dominates and 
  neglected the DM pair production process $f \bar{f} \to \bar{\zeta} \zeta$ from the thermal plasma.

For fixed values of $g_{BL}$, $g_\zeta$ and $m_\zeta$, we numerically solve Eq.~(\ref{BoltzmannDM}) from $x_{RH} \ll 1$. 
In Fig.~\ref{fig:Y2}, we show the yield of the Dirac DM particle as a function of $x=m_\zeta/T$ for $m_\zeta=30$ GeV (solid line),
  along with the yield of $Z_{BL}$ (dashed line). 
Here, we have taken $g_{BL}=5.0 \times 10^{-10}$ and $g_\zeta=6.3 \times 10^{-4}$. 
We can see the result similar to that in Fig.~\ref{fig:Y}. 
As we can understand from Eq.~(\ref{BoltzmannBLY}) and Eq.~(\ref{Eq:App4}), 
  $Y(\infty) \sim Y(x=1) \propto g_\zeta^4 \, g_{BL}^4/m_\zeta^2$.  
We find that the observed DM relic density of $\Omega_{DM} h^2=0.12$ is reproduced when 
 \begin{eqnarray}
 g_\zeta^2 \, g_{BL}^2  \simeq 8.2 \times 10^{-24}  \, \left( \frac{m_\zeta}{2.5 \, {\rm TeV}} \right). 
 \label{sfi}
\end{eqnarray}  
Comparing this result with Eq.~(\ref{case-b}), we conclude that the sequential freeze-in 
  dominantly produces the DM particles for $m_\zeta < 2.5$ TeV, in case (B). 
For $m_\zeta=30$ GeV, our result is displayed in Fig.~\ref{Y1}. 
The plots show cusps at $g_{BL} \simeq 1.5 \times 10^{-7}$, which is the boundary value 
  to separate case (A) and case (B). 
To simplify our analysis, we have calculated the two cases separately by considering only the dominant process 
  in each case.   
Because of this simplification, the cusps appear in our results, and they will be smoothed away 
  if we take all terms into account in the Boltzmann equations.

Thus to summarize, for the freeze-in scenario, there are the following constraints on parameters to reproduce the observed DM relic density depending on the ranges of  $B-L$ gauge coupling $g_{BL}$ and dark matter mass $m_\zeta$. 
Case (A): This constraint applies for  the parameter region $g_{BL}\geq 2.7 \times 10^{-8}\sqrt{m_\zeta[{\rm GeV}]}$ 
  where one has $g^2_\zeta \, g^2_{BL} + \frac{0.82}{1.2} \, g^4_\zeta \simeq 8.2 \times 10^{-24}$
  to reproduce $\Omega_{DM} \, h^2 =0.12$. 
Case (B):  for  $g_{BL} < 2.7 \times 10^{-8} \sqrt{m_\zeta[{\rm GeV}]}$, there are two separate constraints depending on $m_\zeta$. Case (B1): for $m_\zeta \lesssim 2.5$ TeV, we find
$g_\zeta^2 \, g_{BL}^2  \simeq 8.2 \times 10^{-24}  \, \left( \frac{m_\zeta}{2.5 \, {\rm TeV}} \right)$  and in case (B2): for  $m_\zeta \gtrsim 1.5$ TeV, we find
$g_\zeta^2 \, g_{BL}^2  \simeq 8.2 \times 10^{-24}$.
%

%%%%%%%%%%%%%%%%%%%%%%%%%%%%%%%
 \begin{figure}[t]
  \centering
  \includegraphics[width=0.7\linewidth]{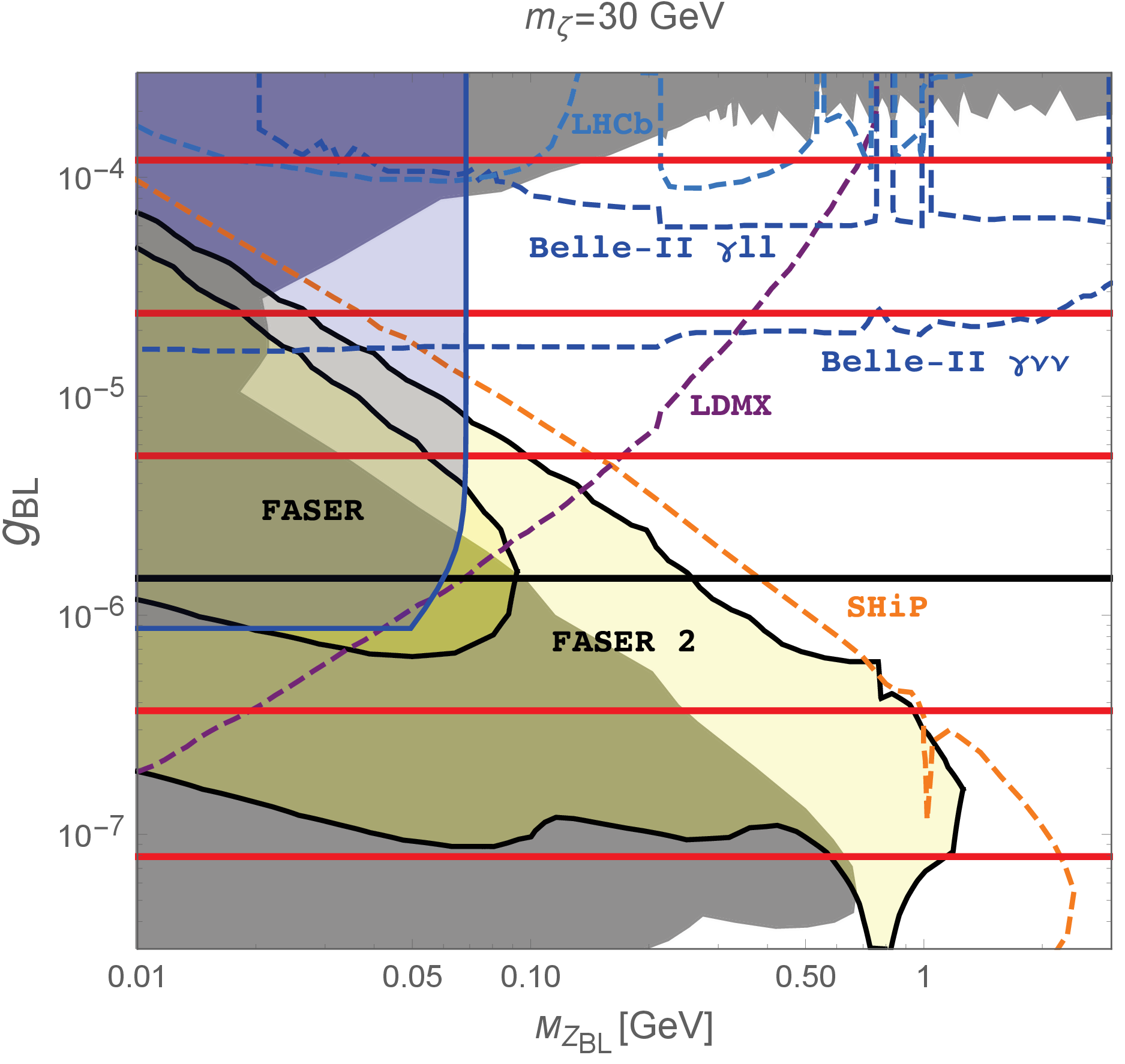}
  \caption{
The various horizontal lines, along which $\Omega_{DM} \, h^2=0.12$ is reproduced, 
   show the results for various $Q$ values: 
   $Q=2 \times 10^{-4}$, $5 \times 10^{-3}$, $0.1$, $1.01$ (black line), $5$, and $50$ from top to bottom. 
We go vertically up as $Q$ decreases (see Eqs.~(\ref{caseA}) and (\ref{sfi})). 
Here, we have chosen $m_\zeta = 30 $ GeV. 
Reaches of the various experiments are shown in different color lines. 
FASER and FASER 2 in solid black lines. 
Orange dashed line is for SHiP~\cite{SHiP},  purple dashed line for LDMX~\cite{LDMX}, 
 dark-blue dashed lines for Belle II \cite{B2}, and light-blue dashed lines for LHCb \cite{LHCb1, LHCb2}.
The region to the left of the solid blue line is excluded by the XENON1T results. 
The line is vertical because $g_{BL} \, g_\zeta$ is almost constant for $g_{BL} \gtrsim 10^{-6}$ 
 (see the right panel in Fig.~\ref{Y1}) in Eq.~(\ref{caseA}). 
For $M_{Z_{BL}} \lesssim 50$ MeV, $\sigma_{SI}$ becomes independent of $M_{Z_{BL}}$ 
  \cite{DelNobile:2015uua, DelNobile:2015bqo, Panci:2014gga, Li:2014vza},  
  the XENON1T bound is satisfied for $g_{BL} \, g_\zeta \lesssim 1.5\times 10^{-12}$. 
This means that the XENON1T constraint is always satisfied for $g_{BL} \lesssim 8.9\times 10^{-7}$ 
  in our scenario. 
}
\label{fig:7}
\end{figure}
%%%%%%%%%%%%%%%%%%%%%

%%%%%%%%%%%%%%%%%%%%%%%%%%%%%%%%%
\subsection{Possible laboratory probes of the freeze-in case}
%%%%%%%%%%%%%%%%%%%%%%%%%%%%%%%%%
We now discuss possible probes of  the freeze-in scenario in the laboratory. 
There are several experiments that can probe various parameter ranges of the model. 
This is shown in Fig.~\ref{fig:7}. 
The relevant experiments are those at the ones attempting to extend lifetime frontier of various new weakly coupled
  beyond the SM particles. They typically look for displaced vertices. 
The experiments are FASER and SHiP at the LHC; Belle II andLHCb
  as well as LDMX experiment proposed to search for weakly coupled light DM particles. 

The planned FASER detector~\cite{faser} at the LHC will probe the low $M_{Z_{BL}}$ ($\leq 1-2$ GeV) and low $g_{BL}$ region of the theory. % is the displaced vertex searches at the planned FASER experiment~\cite{faser} at the  LHC. 
 This is a detector which will be installed in a tunnel near the ATLAS detector about 480 meters away to look for displaced vertices with charged particles from %weakly interacting particles 
long-lived charge-neutral particles produced at the primary LHC vertex. 
In the very low $g_{BL}$ range, our model falls into this category since due to low $g_{BL}$ and low mass $M_{Z_{BL}}$, 
   the distance travelled by a highly boosted $Z_{BL}$ before decaying is given by 
   $c\tau\sim \frac{12\pi E_{Z_{BL}}}{g^2_{BL}M_{Z_{BL}}^2}$ 
   and experiments such as FASER searching for displaced vertices can  give useful constraints.

In Fig.~\ref{fig:7}, the horizontal solid lines correspond to the results for the various $B-L$ charges of the DM particle, 
   $Q=2 \times 10^{-4}$, $5 \times 10^{-3}$, $0.1$, $1.01$ (black line), $5$, and $50$ from top to bottom. 
Along the horizontal lines, $\Omega_{DM} \, h^2=0.12$ is satisfied. 
Various planned and proposed experiments and their search reaches are indicated 
  (see Ref.~\cite{faser} for details) and the current excluded region is gray-shaded \cite{Bauer:2018onh}.  
The blue shaded region at top-left corner is excluded by the XENON1T results. 
As discussed in Sec.~\ref{sec:3.2}, $\sigma_{SI}$ becomes constant for $M_{Z_{BL}} \lesssim 50$ MeV, 
   and we find that the XENON1T bound is satisfied for $g_{BL} \lesssim 8.9 \times 10^{-7}$
   for any values of $M_{Z_{BL}}$.  
Even for the freeze-in case, the direct DM detection experiments provide very severe constraints 
  and exclude a part of the open window.    
 From Fig.~\ref{fig:7}, we see that various Lifetime Frontier experiments in the near future
  can test our freeze-in scenario.

\subsection{Astrophysical and BBN constraints on low mass $Z_{BL}$}

If $Z_{BL}$ mass is less than 100 MeV, it can be produced from $e^+e^-$ and $\nu \bar\nu$ collisions in the supernova, whose core temperature is believed to be 30 MeV. To avoid any constraints on $g_{BL}$ from energy loss considerations of SN 1987A, we stay above $Z_{BL}$ mass of 200 MeV. %There are then two kinds of constraints on  $g_{BL}$ that can be derived from the SN1987A observations \cite{SN1987A1, SN1987A2}. The first kind of constraint arises when the $Z_{BL}$ escapes the SN taking energy away leading to conflicts with the observed energy emitted in neutrinos i.e. $\sim 5\times 10^{53}$ erg/sec. The second kind of bound arises, if $Z_{BL}$ mass is larger than an MeV, it can decay in principle to an $e^+e^-$ pair inside the supernova leading to earlier X-ray and light signals than the three hour time lapse which was observed~\cite{kazanas, gustavo}. We discuss these bounds now.

%%%%%%%%%%%%%%%%%%%%%%%%%%%%
Coming to constraints from Big Bang nucleosynthesis,  we assume that the RHNs required for anomaly cancellation acquire heavy Majorana mass ($M_{N_R}\geq 100$ GeV or more) so that the only new degree of freedom we have to consider at the epoch of BBN are the three modes of the vector boson $Z_{BL}$ (two transverse and one longitudinal). We assume $M_{Z_{BL}}$ to be in the one GeV or lower range but above 200 MeV. 
For the higher mass range, as long as $Z_{BL}$ is in thermal equilibrium, the $Z_{BL}$ density at decoupling is already suppressed enough so that there are no BBN constraints. 

The physics of our considerations in the lower mass range are as follows: if the gauge coupling is large enough that the $Z_{BL}$ is in thermal equilibrium till $T= 1$ MeV, then how much it contributes to the quantity $\Delta N_{eff}$ depends on its mass. If its mass is larger than 10 MeV, its abundance at $T=1$ MeV will be Boltzmann suppressed and its contributions to energy density will be within the current $\Delta N_{eff}$ limits. Since we are interested in the mass range of 200 MeV or more to avoid supernova constraints, we need not worry about the BBN constraints unless the gauge coupling is below $10^{-10}$  GeV in which case it can survive till $T\sim 1$ MeV and affect BBN. The limit of $10^{-10}$ comes from requiring that $\Gamma_{Z_{BL}}\sim H(T=1{\rm MeV})$.

\section{Case (iii): Small $g_{BL}$ and secluded dark sector with $\zeta$ and $Z_{BL}$} 
\label{sec:5}
%%%%%%%%%%%%%%%%%%%%%%%%%%%%%%%%%%%%%%%%%%%%%%%
In this section we briefly comment on two more logical possibilities which arise when $g_{BL} <  2.7 \times 10^{-8}\sqrt{m_{\zeta}[{\rm GeV}]}$
so that the SM particles are decoupled from the $\zeta$ and $Z_{BL}$ sectors. 
There are two possibilities here: case (iiiA) where $g_\zeta$ is large enough so that the DM particle can be in equilibrium with $Z_{BL}$ but not with the SM sector due to small $g_{BL}$, and case (iiiB) where $g_\zeta$ is small so that all three sectors are sequestered. 
Here we comment briefly on how the relic density can arise in both of the cases.

In either of cases (iiiA) and (iiiB), the decay of the inflaton will play a crucial role in building up the DM relic density. 
Assuming the inflaton $\phi$ being a gauge singlet scalar under the SM and $B-L$ gauge groups, 
  we can consider couplings of the inflaton with particles in our model 
  such as $c_H \, \phi H^\dagger H$,  $c_\zeta \, \phi \bar\zeta \zeta$ and $ c_Z \, \phi {\cal Z}_{BL}^{\mu\nu} {\cal Z}_{BL\, \mu \nu}$, 
  where $H$ is the SM Higgs doublet, ${\cal Z}_{BL}^{\mu\nu} $ is the field strength of $Z_{BL}$, 
  $c_H$ is a coupling with a mass dimension $+1$,  
  $c_\zeta$ is a dimensionless coupling, and $c_Z$ is a coupling with a mass-dimension $-1$. 
After the end of inflation, the inflaton decays to particles through these couplings to reheat the universe 
  and then the Big Bang Hubble era begins. 
Assuming that the inflaton is much heavier than any other particles, the inflaton partial decay widths are calculated as
\begin{eqnarray}
\Gamma_{\phi \to H^\dagger H} &=&  \frac{c_H^2}{8 \pi \, m_\phi},  \nonumber \\
\Gamma_{\phi \to \bar{\zeta} \zeta} &=&  \frac{c_\zeta^2}{8 \pi} m_\phi,   \nonumber \\
\Gamma_{\phi \to Z_{BL} Z_{BL}} &=&  \frac{c_Z^2}{4 \pi}  m_\phi^3.
\end{eqnarray}
We consider that the inflaton mainly decays to the Higgs doublets and the reheating temperature
(of the SM particle plasma)  is estimated by $\Gamma_{\phi \to H^\dagger H} \simeq H(T_{RH})$, 
so that
\begin{eqnarray}
  T_{RH} \simeq \sqrt{\Gamma_{\phi \to H^\dagger H} \, M_P} \sim c_H \sqrt{\frac{M_P}{m_\phi}}.
\end{eqnarray}

For case (ii) in Sec.~\ref{sec:4}, we implicitly assumed that the branching raito of the inflaton decay into the DM particles 
  is negligibly small so that we employed the initial condition $Y(x_{RH})=0$ in solving the Boltzmann equation. 
Here in case (iii), we are considering the case where the inflaton branching ratio into the ``dark sector'' with $\zeta$ and $Z_{BL}$ 
  is not negligible. There are then two possible cases.
  
For case (iiiA), the early universe after reheating consists of two separate plasmas: one is the thermal plasma 
  of the SM particles and the other is the plasma of the hidden sector, where $\zeta$ and $Z_{BL}$ are in thermal equilibrium. 
Note that the formula to evaluate the reheating temperature, $\Gamma_{\phi \to H^\dagger H} \simeq H(T_{RH})$,  
  means that the inflaton energy at its lifetime is transmitted to the SM particles plasma. 
Thus, we estimate the reheating temperature of the dark sector by 
\begin{eqnarray}
  T_{RH}^{\rm{dark~sector}} \simeq \sqrt{\Gamma_{\phi \to H^\dagger H} \, M_P} 
  \times \sqrt{BR(\phi \to \bar{\zeta} \zeta)+BR(\phi \to Z_{BL} Z_{BL})}, 
\end{eqnarray}
where $BR(\phi \to \bar{\zeta} \zeta)$ and $BR(\phi \to Z_{BL} Z_{BL})$ are the inflaton branching ratios 
  to $\bar{\zeta} \zeta$ and $Z_{BL} Z_{BL}$, respectively. 
Although the temperatures of the SM sector and the dark sector are not the same, 
  unless the branching ratio is extremely small, 
  the evaluation of the DM relic density is similar to case (i) discussed in Sec.~\ref{sec:3}. 

For case (iiiB) on the other hand, all three sectors are sequestered. 
The energy density of the dark matter sector at the reheating is estimated by
\begin{eqnarray}
  \rho_\zeta \simeq BR(\phi \to \bar{\zeta} \zeta) \times \rho_{rad}(T_{RH}), 
\end{eqnarray}
where $\rho_{rad} = \frac{\pi^2}{30} g_* T_{RH}^4$ is the energy density of the SM particle plasma. 
For a given $T_{RH}$ value, we may adjust the inflaton branching ratio 
  into a pair of DM particles to reproduce the observed DM relic density.

As a final comment, we note that one may identify the inflaton field with the $B-L$ breaking Higgs boson ($\Delta$). 
In this case, we consider couplings of the inflaton such as 
  $\lambda_{mix}  \, \Delta^\dagger \Delta  H^\dagger H$,  
  $ \frac{c_\Delta}{M_P} \, \Delta^\dagger \Delta \bar\zeta \zeta$ and 
  $ g_{BL}^2  \Delta^\dagger \Delta  Z_{BL}^{\mu}  Z_{BL\, \mu}$, 
where $\lambda_{mix}$ and $c_\Delta$ are dimensionless coupling constants.  
We can apply the above discussion by the replacements:  
  $\phi \to \sigma$,  $c_H \to \lambda_{mix} \, v_{BL}$, $c_\zeta \to v_{BL}/M_P$ and $c_Z \to g^2 v_{BL}$.

%%%%%%%%%%%%%%%%%%%%%
\section{Summary and conclusions} 
\label{sec:6}
%%%%%%%%%%%%%%%%%%%%%
In summary, we have considered an extension of the standard model with the gauged $U(1)_{B-L}$ symmetry and  a Dirac fermion witharbitrary  $B-L$ charge which plays the role of dark matter. 
The $B-L$ symmetry is broken by a $B-L=2$ Higgs field so that $Z_{BL}$ picks up a mass and it leads to the seesaw mechanism for neutrino masses. This provides  a unified picture of neutrinos and dark matter.
Ignoring the mixings of $Z_{BL}$ with SM gauge bosons, we show that  in the weakly coupled $B-L$ gauge boson case 
  there are constraints on the gauge couplings $g_{BL}$ of SM fermions and $g_\zeta$ of dark matter  as well as the masses of the dark matter 
  and $M_{Z_{BL}}$ 
  from different observations such as Fermi-LAT,  CMB, $\Omega_{DM} h^2$ 
  and direct dark matter detection experiments for the case when the dark matter is a thermal freeze-out type.  
We also point out that for even weaker gauge couplings where the dark matter relic density arises via the freeze-in mechanism, 
  there are constraints on the above couplings from the observed dark matter relic density as well as from the supernova 1987A observations. 
We note that parts of the freeze-in parameter range of the model can be tested 
  in the FASER experiment being planned at the LHC and other ``Lifetime Frontier'' experiments.

%%%%%%%%%%%%%%%%
\section*{Acknowledgement} 
%%%%%%%%%%%%%%%%
N.O. would like to thank the Maryland Center for Fundamental Physics for hospitality during his visit. 
%The authors would like to thank the referees for very useful comments which helped to improve the paper.
The work of R.N.M. is supported by the National Science Foundation grant No.~PHY1620074 and PHY-1914631 and
the work of N.O. is supported by the US Department of Energy grant No.~DE-SC0012447. 
\\

\noindent{\bf Note added in proof}\\ 
After this work was put in the arXiv, the paper arXiv:1908.09834~\cite{felix} with a similar study was brought to our attention.

%%%%%%%%%%%%%%%%%%%%%
%\appendix
\section*{Appendix} 
%%%%%%%%%%%%%%%%%%%%%
In this appendix, we list the formulas that we have used in our analysis. 

For the annihilation process of $\zeta \bar{\zeta} \to Z_{BL} \to f\bar{f}$, 
the cross section times relative velocity is given by 
\begin{eqnarray}
\sigma v =  \frac{g^2_{\zeta} \, g_{BL}^2}{6 \pi s}  \sum_{f} N_{c}^{f} \, Q_f^2 \,
\frac{
\left( s+ 2 m_{\zeta}^2 \right)
\left( s+ 2 m_f^2 \right) 
}{(s-M_{Z_{BL}}^2)^2}
\sqrt{1-\frac{4 m_f^2}{s}}  ,
\label{Eq:App1}
\end{eqnarray}
where $f$ denotes a SM fermion with mass of $m_f$, $Q_f$ is its $B-L$ charge, 
and $N_{c}^f$ is the color number in the final state of a SM fermion:
$N_{c}^{f}=3$ for a quark, $N_{c}^{f}=1$ for a charged lepton, $N_{c}^{f}=1/2$ for a SM neutrino $\left(m_{f} \rightarrow 0\right)$.
Since we are interested in the case of $m_\zeta > M_{Z_{BL}}$, 
we have neglected the decay width of the $Z_{BL}$ boson in the above formula. 
In the non-relativistic limit, the cross section formula is simplified to be 
\begin{eqnarray}
\sigma v  \simeq  \frac{g^2_{\zeta} \, g_{BL}^2}{2  \pi}  \sum_{f} N_{c}^{f} \, Q_f^2 \,
\frac{2 m_\zeta^2+m_f^2}{(4 m_\zeta^2-M_{Z_{BL}}^2)^2}
\sqrt{1-\frac{m_f^2}{m_\zeta^2}}  , 
\label{Eq:App2}
\end{eqnarray}
while in the relativistic limit, 
\begin{eqnarray}
\sigma v \simeq  \frac{g^2_{\zeta} \, g_{BL}^2}{6 \pi s}  \sum_{f} N_{c}^{f} \, Q_f^2.
\label{Eq:App3}
\end{eqnarray}

For the annihilation process of $\zeta \bar{\zeta} \to Z_{BL}  Z_{BL}$, 
the cross section times relative velocity is given by 
\begin{eqnarray}
\sigma v &=&  \frac{g^4_{\zeta}}{4 \pi s}  \, \sqrt{1-\frac{4 m_{Z_{BL}}^2}{s}} \nonumber \\
&\times& \left(
-1 - \frac{(2+a^2)^2}{(2-a^2)^2+4 b^2}
+ \frac{6-2 a^2+a^4+12 b^2+4 b^4}{2 b c (1+b^2+c^2)} 
\ln \left[\frac{1+(b+c)^2}{1+(b-c)^2} \right]
\right),
\label{Eq:App4}
\end{eqnarray}
where $a=\frac{M_{Z_{BL}}}{m_\zeta}$, $b=\sqrt{\frac{s}{4 m_\zeta^2}-1}$, and  
 $c=\sqrt{\frac{s}{4 m_\zeta^2}-a^2}$. 
In the non-relativistic limit, this cross section formula is simplified to be 
\begin{eqnarray}
\sigma v \simeq \frac{g_{\zeta}^{4}}{16\pi m_{\zeta}^{\, 2}} \left(1-\frac{M_{Z_{BL}}^{2}}{m_{\zeta}^{\, 2}} \right)^{3/2}
\left(1-\frac{M_{Z_{BL}}^{2}}{2 m_{\zeta}^{\, 2}} \right)^{-2}, 
\label{Eq:App5}
\end{eqnarray}
while in the relativistic limit, 
\begin{eqnarray}
\sigma v \simeq \frac{g_{\zeta}^{4}}{4 \pi s} \left( \ln \left[\frac{s}{m_\zeta^2} \right] -1 \right). 
\label{Eq:App6}
\end{eqnarray}
%

%\newpage
%%%%%%%%%%%%%%%%%
%\section*{References} 
%%%%%%%%%%%%%%%%%


\begin{thebibliography}{99}
%%%%%%%%%%%%%%%%%
 
\bibitem{marshak1}  R.~E.~Marshak and R.~N.~Mohapatra,
  %``Quark - Lepton Symmetry and B-L as the U(1) Generator of the Electroweak Symmetry Group,''
  Phys.\ Lett.\  {\bf 91B}, 222 (1980).
 

\bibitem{marshak2} R. N. Mohapatra and R. E. Marshak, Phys. Rev. letters {\bf 44}, 1316  (1980).

\bibitem{Davidson}
 A.~Davidson,
  %``$B^-$l as the Fourth Color, Quark - Lepton Correspondence, and Natural Masslessness of Neutrinos Within a Generalized Ws Model,''
  Phys.\ Rev.\ D {\bf 20}, 776 (1979). 


\bibitem{BL0} W.~Buchmuller, C.~Greub and P.~Minkowski,
  %``Neutrino masses, neutral vector bosons and the scale of B-L breaking,''
  Phys.\ Lett.\ B {\bf 267}, 395 (1991).
  

\bibitem{khalil1} S.~Khalil,
  %``Low scale $B$ - L extension of the Standard Model at the LHC,''
  J.\ Phys.\ G {\bf 35}, 055001 (2008).
  


\bibitem{BL1} L.~Basso,
  %``Phenomenology of the minimal B-L extension of the Standard Model at the LHC,''
  arXiv:1106.4462 [hep-ph].
  
\bibitem{BL1a}
  L.~Basso, A.~Belyaev, S.~Moretti and C.~H.~Shepherd-Themistocleous,
  %``Phenomenology of the minimal B-L extension of the Standard model: Z' and neutrinos,''
  Phys.\ Rev.\ D {\bf 80}, 055030 (2009). 
%  doi:10.1103/PhysRevD.80.055030
%  [arXiv:0812.4313 [hep-ph]].  
  
  
  \bibitem{BL3} 
  S.~Iso, N.~Okada and Y.~Orikasa,
  %``Classically conformal $B^-$ L extended Standard Model,''
  Phys.\ Lett.\ B {\bf 676}, 81 (2009); 
%  doi:10.1016/j.physletb.2009.04.046
%  [arXiv:0902.4050 [hep-ph]]. 
%
%\cite{Iso:2009nw}
%\bibitem{Iso:2009nw} 
%  S.~Iso, N.~Okada and Y.~Orikasa,
  %``The minimal B-L model naturally realized at TeV scale,''
  Phys.\ Rev.\ D {\bf 80}, 115007 (2009).  
%  doi:10.1103/PhysRevD.80.115007
%  [arXiv:0909.0128 [hep-ph]].



\bibitem{BL2}  A.~A.~Abdelalim, A.~Hammad and S.~Khalil,
  %``B-L heavy neutrinos and neutral gauge boson Z? at the LHC,''
  Phys.\ Rev.\ D {\bf 90}, no. 11, 115015 (2014)
  

  
  \bibitem{BL4} A.~Biswas, S.~Choubey and S.~Khan,
  %``Inverse seesaw and dark matter in a gauged B-L extension with flavour symmetry,''
  JHEP {\bf 1808}, 062 (2018)
  
  

\bibitem{seesaw1} P. Minkowski, Phys. Lett. B {\bf 67}, 421 (1977).

\bibitem{seesaw2} R. N. Mohapatra and G. Senjanovi\'{c}, Phys. Rev. Lett. {\bf 44}, 912 (1980).

\bibitem{seesaw3} T. Yanagida, Conf.  Proc.  C {\bf 7902131},  95  (1979).

\bibitem{seesaw4} M. Gell-Mann, P. Ramond and R. Slansky, Conf. Proc. C {\bf 790927}, 315 (1979) [arXiv:1306.4669 [hep-th]].

\bibitem{seesaw5} S.~L.~Glashow, NATO Sci. Ser. B {\bf 61}, 687 (1980).



\bibitem{garv} G.~Chauhan, P.~S.~B.~Dev, R.~N.~Mohapatra and Y.~Zhang,
  %``Perturbativity constraints on $U(1)_{B-L}$ and left-right models and implications for heavy gauge boson searches,''
  JHEP {\bf 1901}, 208 (2019).
  
\bibitem{FileviezPerez:2019cyn} 
  P.~Fileviez Perez, C.~Murgui and A.~D.~Plascencia,
  %``Neutrino-Dark Matter Connections in Gauge Theories,''
  Phys.\ Rev.\ D {\bf 100}, 035041 (2019). 
%  doi:10.1103/PhysRevD.100.035041
%  [arXiv:1905.06344 [hep-ph]].  


\bibitem{Gu:2019ohx} 
  P.~H.~Gu,
  %``Dirac seesaw accompanied by Dirac fermionic dark matter,''
  arXiv:1907.10018 [hep-ph].
  
  \bibitem{burgess} C. P. Burgess, J. P. Conlon, L. Y. Hung, C. H. Kom, A. Maharana and F. Quevedo, JHEP{\bf  0807}, 073 (2008).  % [arXiv:0805.4037[hep-th]]. 
  
  \bibitem{hall} L.~J.~Hall, K.~Jedamzik, J.~March-Russell and S.~M.~West,
  %``Freeze-In Production of FIMP Dark Matter,''
  JHEP {\bf 1003}, 080 (2010).
  
  \bibitem{bernal}  N.~Bernal, M.~Heikinheimo, T.~Tenkanen, K.~Tuominen and V.~Vaskonen,
  %``The Dawn of FIMP Dark Matter: A Review of Models and Constraints,''
  Int.\ J.\ Mod.\ Phys.\ A {\bf 32}, no. 27, 1730023 (2017). 
  %doi:10.1142/S0217751X1730023X
 % [arXiv:1706.07442 [hep-ph]].
 
  \bibitem{hambye} T.~Hambye, M.~H.~G.~Tytgat, J.~Vandecasteele and L.~Vanderheyden,
  %``Dark matter direct detection is testing freeze-in,''
  Phys.\ Rev.\ D {\bf 98}, no. 7, 075017 (2018).
  %doi:10.1103/PhysRevD.98.075017
%  [arXiv:1807.05022 [hep-ph]].

\bibitem{chu} X.~Chu, T.~Hambye and M.~H.~G.~Tytgat,
  %``The Four Basic Ways of Creating Dark Matter Through a Portal,''
  JCAP {\bf 1205}, 034 (2012).  
  %doi:10.1088/1475-7516/2012/05/034
%  [arXiv:1112.0493 [hep-ph]].


\bibitem{Kaneta}
  K.~Kaneta, Z.~Kang and H.~S.~Lee,
  %``Right-handed neutrino dark matter under the $B − L$ gauge interaction,''
  JHEP {\bf 1702}, 031 (2017). 
%  doi:10.1007/JHEP02(2017)031
%  [arXiv:1606.09317 [hep-ph]].
  
  
  
\bibitem{fermi} A. Albert et al. 1611.03184, [Astro-ph.HE]
  
  
  
  \bibitem{bauer} M.~Bauer, P.~Foldenauer and J.~Jaeckel,
  %``Hunting All the Hidden Photons,''
  JHEP {\bf 1807}, 094 (2018).
  
  \bibitem{biswas} A.~Biswas and A.~Gupta,
  %``Freeze-in Production of Sterile Neutrino Dark Matter in U(1)$_{\rm B-L}$ Model,''
  JCAP {\bf 1609}, 044 (2016)
 
  
  \bibitem{heeck} J.~Heeck,
  %``Unbroken B - L symmetry,''
  Phys.\ Lett.\ B {\bf 739}, 256 (2014).
  

  
  \bibitem{lindner} M.~Dutra, M.~Lindner, S.~Profumo, F.~S.~Queiroz, W.~Rodejohann and C.~Siqueira,
  %``MeV Dark Matter Complementarity and the Dark Photon Portal,''
  JCAP {\bf 1803}, 037 (2018).
  
   \bibitem{cirelli} M.~Cirelli, P.~Panci, K.~Petraki, F.~Sala and M.~Taoso,
  %``Dark Matter's secret liaisons: phenomenology of a dark U(1) sector with bound states,''
  JCAP {\bf 1705}, 036 (2017).
 
 
  
  \bibitem{faser1} J.~L.~Feng, I.~Galon, F.~Kling and S.~Trojanowski,
  %``ForwArd Search ExpeRiment at the LHC,''
  Phys.\ Rev.\ D {\bf 97}, no. 3, 035001 (2018)
  %doi:10.1103/PhysRevD.97.035001
  [arXiv:1708.09389 [hep-ph]].
 
\bibitem{nobu2}  R.~N.~Mohapatra and N.~Okada,
  %``Freeze-in Dark Matter from a Minimal B-L Model and Possible Grand Unification,''
   Phys.\ Rev.\ D {\bf 101}, 115022 (2020).
  
  \bibitem{LEP_rev} Electroweak [LEP and ALEPH and DELPHI and L3 and OPAL Collaborations and LEP Electroweak Working Group and SLD Electroweak Group and SLD Heavy Flavor Group], hep-ex/0312023.
 

\bibitem{ATLbound}
M.~Aaboud {\it et al.} [ATLAS Collaboration],
  %``Search for new high-mass phenomena in the dilepton final state using 36 fb$^{−1}$ of proton-proton collision data at $ \sqrt{s}=13 $ TeV with the ATLAS detector,''
  JHEP {\bf 1710}, 182 (2017). 
  
  \bibitem{Lang} Paul Langacker, Rev. Mod. Phys. {\bf 81}, 1199 (2009).
    
  
  
\bibitem{farinaldo}  A.~Alves, A.~Berlin, S.~Profumo and F.~S.~Queiroz,
  %``Dirac-fermionic dark matter in U(1)$_{X}$ models,''
  JHEP {\bf 1510}, 076 (2015);  Phys.\ Rev.\ D {\bf 92}, no. 8, 083004 (2015).


\bibitem{review1}
 E.~W.~Kolb and M.~S.~Turner,
  %``The Early Universe,''
  Front.\ Phys.\  {\bf 69}, 1 (1990).

 \bibitem{review2} G.~Bertone, D.~Hooper and J.~Silk,
  %``Particle dark matter: Evidence, candidates and constraints,''
  Phys.\ Rept.\  {\bf 405}, 279 (2005)
 
\bibitem{Planck2019}
 N.~Aghanim {\it et al.} [Planck Collaboration],
  %``Planck 2018 results. VI. Cosmological parameters,''
  arXiv:1807.06209 [astro-ph.CO].
    

\bibitem{Xenon1T-2018} 
  E.~Aprile {\it et al.} [XENON Collaboration],
  %``Dark Matter Search Results from a One Ton-Year Exposure of XENON1T,''
  Phys.\ Rev.\ Lett.\  {\bf 121}, no. 11, 111302 (2018). 
%  doi:10.1103/PhysRevLett.121.111302
%  [arXiv:1805.12562 [astro-ph.CO]].


%DarkSide-50 
 \bibitem{DarkSide-50} 
  P.~Agnes {\it et al.} [DarkSide Collaboration],
  %``Low-Mass Dark Matter Search with the DarkSide-50 Experiment,''
  Phys.\ Rev.\ Lett.\  {\bf 121}, no. 8, 081307 (2018).
%  doi:10.1103/PhysRevLett.121.081307
%  [arXiv:1802.06994 [astro-ph.HE]].
 
  
% LIUX 
 \bibitem{LUX-2019} 
  D.~S.~Akerib {\it et al.} [LUX Collaboration],
  %``Extending light WIMP searches to single scintillation photons in LUX,''
  arXiv:1907.06272 [astro-ph.CO].
 
 %PandaX-II  
 \bibitem{PandaX-II} 
  X.~Cui {\it et al.} [PandaX-II Collaboration],
  %``Dark Matter Results From 54-Ton-Day Exposure of PandaX-II Experiment,''
  Phys.\ Rev.\ Lett.\  {\bf 119}, no. 18, 181302 (2017). 
%  doi:10.1103/PhysRevLett.119.181302
%  [arXiv:1708.06917 [astro-ph.CO]].
 
 

%\cite{DelNobile:2015uua}
\bibitem{DelNobile:2015uua} 
  E.~Del Nobile, M.~Kaplinghat and H.~B.~Yu,
  %``Direct Detection Signatures of Self-Interacting Dark Matter with a Light Mediator,''
  JCAP {\bf 1510}, 055 (2015)
  doi:10.1088/1475-7516/2015/10/055
  [arXiv:1507.04007 [hep-ph]]. 
  
  %\cite{DelNobile:2015bqo}
\bibitem{DelNobile:2015bqo} 
  E.~Del Nobile, M.~Nardecchia and P.~Panci,
  %``Millicharge or Decay: A Critical Take on Minimal Dark Matter,''
  JCAP {\bf 1604}, 048 (2016). 
%  doi:10.1088/1475-7516/2016/04/048
%  [arXiv:1512.05353 [hep-ph]].
    
%\cite{Panci:2014gga}
\bibitem{Panci:2014gga} 
  P.~Panci,
  %``New Directions in Direct Dark Matter Searches,''
  Adv.\ High Energy Phys.\  {\bf 2014}, 681312 (2014). 
%  doi:10.1155/2014/681312
%  [arXiv:1402.1507 [hep-ph]].   
 
%\cite{Li:2014vza}
\bibitem{Li:2014vza} 
  T.~Li, S.~Miao and Y.~F.~Zhou,
  %``Light mediators in dark matter direct detections,''
  JCAP {\bf 1503}, 032 (2015). 
%   doi:10.1088/1475-7516/2015/03/032
%  [arXiv:1412.6220 [hep-ph]]. 
    


\bibitem{walia} T.~Bringmann, F.~Kahlhoefer, K.~Schmidt-Hoberg and P.~Walia,
  %``Strong constraints on self-interacting dark matter with light mediators,''
  Phys.\ Rev.\ Lett.\  {\bf 118}, no. 14, 141802 (2017)

%\bibitem{cirelli}  M.~Cirelli, P.~Panci, K.~Petraki, F.~Sala and M.~Taoso,
  %``Dark Matter's secret liaisons: phenomenology of a dark U(1) sector with bound states,''
%  JCAP {\bf 1705}, 036 (2017);
   
 
 
 \bibitem{Hambye:2019dwd} 
  T.~Hambye, M.~H.~G.~Tytgat, J.~Vandecasteele and L.~Vanderheyden,
  %``Dark matter from dark photons: a taxonomy of dark matter production,''
  Phys.\ Rev.\ D {\bf 100}, no. 9, 095018 (2019)
%  doi:10.1103/PhysRevD.100.095018
  [arXiv:1908.09864 [hep-ph]].

  
  \bibitem{faser} A.~Ariga {\it et al.} [FASER Collaboration],
  %``FASER?s physics reach for long-lived particles,''
  Phys.\ Rev.\ D {\bf 99}, no. 9, 095011 (2019).
  
  
\bibitem{SHiP}
  S.~Alekhin {\it et al.},
  %``A facility to Search for Hidden Particles at the CERN SPS: the SHiP physics case,''
  Rept.\ Prog.\ Phys.\  {\bf 79}, no. 12, 124201 (2016).
%  doi:10.1088/0034-4885/79/12/124201
%  [arXiv:1504.04855 [hep-ph]]. 
  
    
  \bibitem{LDMX}  T.~Akesson {\it et al.} [LDMX Collaboration],
  %``Light Dark Matter eXperiment (LDMX),''
  arXiv:1808.05219 [hep-ex].
  
\bibitem{B2}  
M.~J.~Dolan, T.~Ferber, C.~Hearty, F.~Kahlhoefer, and K.~Schmidt-Hoberg, 
   %“Revised constraints and Belle II sensitivity for visible and invisible axion-like particles,” 
 JHEP {\bf 12}, 094 (2017). 

\bibitem{LHCb1}
 P.~Ilten, J.~Thaler, M.~Williams, and W.~Xue, 
 %“Dark photons from charm mesons at LHCb,”    
 Phys.\ Rev.\ D92, no. 11, 115017 (2015). 
 

\bibitem{LHCb2}
P.~Ilten, Y.~Soreq, J.~Thaler, M.~Williams, and W.~Xue, 
%“Proposed Inclusive Dark Photon Search at LHCb,” 
Phys.\ Rev.\ Lett.\  {\bf 116} no. 25, 251803 (2016). 


\bibitem{Bauer:2018onh} 
  M.~Bauer, P.~Foldenauer and J.~Jaeckel,
  %``Hunting All the Hidden Photons,''
  JHEP {\bf 1807}, 094 (2018)
  [JHEP {\bf 2018}, 094 (2020)]. 
%  doi:10.1007/JHEP07(2018)094
%  [arXiv:1803.05466 [hep-ph]].
  
  
%\bibitem{SN1987A1} 
%  K.~Hirata {\it et al.} [Kamiokande-II Collaboration],
%  %``Observation of a Neutrino Burst from the Supernova SN 1987a,''
%  Phys.\ Rev.\ Lett.\  {\bf 58}, 1490 (1987).
%%   doi:10.1103/PhysRevLett.58.1490  
  
%\bibitem{SN1987A2} 
%  R.~M.~Bionta {\it et al.},
%  %``Observation of a Neutrino Burst in Coincidence with Supernova SN 1987a in the Large Magellanic Cloud,''
%  Phys.\ Rev.\ Lett.\  {\bf 58}, 1494 (1987).
%  doi:10.1103/PhysRevLett.58.1494


%  \bibitem{kazanas} D.~Kazanas, R.~N.~Mohapatra, S.~Nussinov, V.~L.~Teplitz and Y.~Zhang,
%  %``Supernova Bounds on the Dark Photon Using its Electromagnetic Decay,''
%  Nucl.\ Phys.\ B {\bf 890}, 17 (2014). 
  
%  \bibitem{gustavo} W.~DeRocco, P.~W.~Graham, D.~Kasen, G.~Marques-Tavares and S.~Rajendran,
%  %``Observable signatures of dark photons from supernovae,''
%  JHEP {\bf 1902}, 171 (2019)
%  doi:10.1007/JHEP02(2019)171
%  [arXiv:1901.08596 [hep-ph]].
 
 
\bibitem{felix} S.~Heeba and F.~Kahlhoefer,
  %``Probing the freeze-in mechanism in dark matter models with $U(1)^\prime$ gauge extensions,''
  arXiv:1908.09834 [hep-ph].
 
 
 
 
%\bibvitem{SK2}


\end{thebibliography}
\end{document}